 \documentclass[twocolumn,showpacs]{revtex4}

\usepackage[latin1]{inputenc}
\usepackage{graphicx}
\usepackage{dcolumn}
\usepackage{amsmath}
\usepackage[none]{hyphenat}

\makeatletter
\def\btt#1{\texttt{\@backslashchar#1}}%
\DeclareRobustCommand\bblash{\btt{\@backslashchar}}%
\makeatother

\begin{document}

  
 \title{Physics of Automated-Driving Vehicular Traffic }

\author{Boris S. Kerner$^1$}

 \affiliation{$^1$
Physics of Transport and Traffic, University of Duisburg-Essen,
47048 Duisburg, Germany}

\pacs{89.40.-a, 47.54.-r, 64.60.Cn, 05.65.+b}

\begin{abstract} 
We have found that a   variety of  
phase transitions   occurring between  
three traffic phases (free flow (F), synchronized flow (S), and wide moving jam (J))     determine the spatiotemporal dynamics of
  traffic consisting of 100$\%$
 automated-driving vehicles moving  on a two-lane road with an on-ramp bottleneck.  This means that three-phase traffic theory is a common framework
for the description of traffic states   independent of whether  human-driving or automated-driving vehicles
  move in vehicular traffic. To prove this, we have studied automated-driving vehicular traffic with the use
	of classical   Helly's model (1959) 
 widely applied for automated  vehicle motion.
  Although  dynamic rules   of the motion of 
  automated-driving vehicles in a road lane are qualitatively different from those of   human-driving vehicles,
we have revealed that a free-flow-to-synchronized-flow transition
(F$\rightarrow$S transition) exhibits the
nucleation nature, which was   observed in
empirical field   data measured in  traffic consisting of 100$\%$
 human-driving vehicles.  The physics of the nucleation nature of the F$\rightarrow$S transition in automated-driving traffic
is associated with a discontinuity in the rate of lane-changing       that causes
  the
	discontinuity in the rate of  over-acceleration.
 This discontinuous character of over-acceleration   leads to
both the existence and self-maintaining of   synchronized flow  at the bottleneck in
automated-driving  vehicular traffic   as well as to  the existence  at any time instant of
 a range of highway capacities between some minimum and maximum capacities. Within the capacity range, an
F$\rightarrow$S transition can be induced; however, when the maximum capacity is exceeded, then
after some time-delay a spontaneous
  F$\rightarrow$S transition  occurs at the bottleneck.
The phases F, S, and J  can coexist each other in space and time. 
 \end{abstract}

\maketitle

\section{Introduction \label{Int_S}}

In traffic of human-driving vehicles, traffic  breakdown that is a transition from free flow to congested traffic occurs mostly at   bottlenecks. Already in 1950s--1960s two classes of   models for traffic  breakdown were introduced:

(i) In the classical Lighthill-Whitham-Richards (LWR) model~\cite{LighthillWhitham,Richards},
 it is assumed that there is 
 a fundamental diagram for traffic flow at a highway
 bottleneck; the maximum flow rate at the fundamental diagram is equal to  highway capacity:
  If the flow rate upstream of a bottleneck exceeds the capacity, traffic breakdown occurs; otherwise, no traffic breakdown can occur
	at the bottleneck (see, e.g.,~\cite{May,Daganzo1997,Manual2010,Roess2014,Ni2015}).

 (ii)   In 1958, Herman, Gazis, Montroll, Potts, Rothery and Chandler from General Motors (GM) Company~\cite{Chandler,GH195910,Gazis1961A10,GH10} as well as by Kometani and Sasaki~\cite{KS1,KS2,KS3,KS4} assumed that traffic breakdown occurs due to  traffic flow instability in vehicular traffic. This  classical traffic instability  was incorporated into a number of traffic flow models (e.g., papers, reviews,
 and books~\cite{Newell1961,Newell1963A,Gipps,NSch1992,Bando1994,Reviews1,Reviews2,Reviews3,Reviews4,Saifuzzaman2015A,Schadschneider_Book,Treiber2013,Ashton1966,GerloughHuber1975,Gazis2002,Barcelo2010,Elefteriadou2014,Roess2014,Ni2015,Kessels2019}).   As found in~\cite{KernerKon1993}, the classic traffic instability leads to a phase transition from free flow (F) to a wide moving jam (J) called an F$\rightarrow$J transition. 

It is commonly assumed that in future  vehicular traffic    automated-driving vehicles [automated vehicle (AV)]
	 will play a decisive role (see, e.g.,~\cite{Ioannou2006,IoannouChien2002A,Levine1966A_Aut,Liang1999A_Aut,Liang2000A_Aut,Swaroop1996A_Aut,Varaiya1993A,Lin2009A,Martinez2007A,Brummelen2018A,fail_Shladover1995A,Rajamani2012A_Aut,Davis2004B9,Davis2014C,Davis2015A_Int1}). Automated-driving is realized through the use of an
   automated  system in a vehicle that controls over the vehicle in traffic flow as well as through the use of cooperative driving realized
	through   vehicle-to-vehicle communication   or/and through vehicle-to-infrastructure communication 
	(see, e.g.,~\cite{Schmitz2004,Maurer2006,Chen2006,Wang2020,Chen2015}). In most
	studies of the effect of automated vehicles on mixed traffic consisting of random distribution of automated-driving and human-driving vehicles (e.g., see~\cite{Dharba1999A,Marsden2001A,Krishnan2001A,VanderWerf2002A,Shrivastava2002A,Kukuchi2003A,BoseIoannou2003A,Suzuki2003A,Zhou2005A,vanArem2006,Kesting2007A,Ngoduy2012A,Papageorgiou2015B,Talebpour2016A,Wang2017A,Mamouei2018A,Sharon2017A,HanAhn2018A,Zhou2017B,Klawtanong2020A,Geng_Zhang2019A,H_B_Zhu2020A,ZhouZhu2020A,Zhou_Ahn2020A,Danjue_Chen2019A,Zijia_Zhong2020A,Fangfang_Zheng2020A,Shuang_Jin2020A,Zhou_Zhu2020A,ZhihongYao2019A,YeYamamoto2020A,ZhuZhang2018A,Wen-Xing2018A,YeYamamoto2018A,Xin_Chang2020A}), motion of human-driving vehicles  is described with the use of the above-mentioned standard traffic flow models~\cite{May,Daganzo1997,Manual2010,LighthillWhitham,Richards,Chandler,GH195910,Gazis1961A10,GH10,KS1,KS2,KS3,KS4,Newell1961,Newell1963A,Gipps,NSch1992,Bando1994,Reviews1,Reviews2,Reviews3,Reviews4,Saifuzzaman2015A,Schadschneider_Book,Treiber2013,Ashton1966,GerloughHuber1975,Gazis2002,Barcelo2010,Elefteriadou2014,Roess2014,Ni2015,Kessels2019}.  

\begin{figure}
\begin{center}
\includegraphics[width = 7 cm]{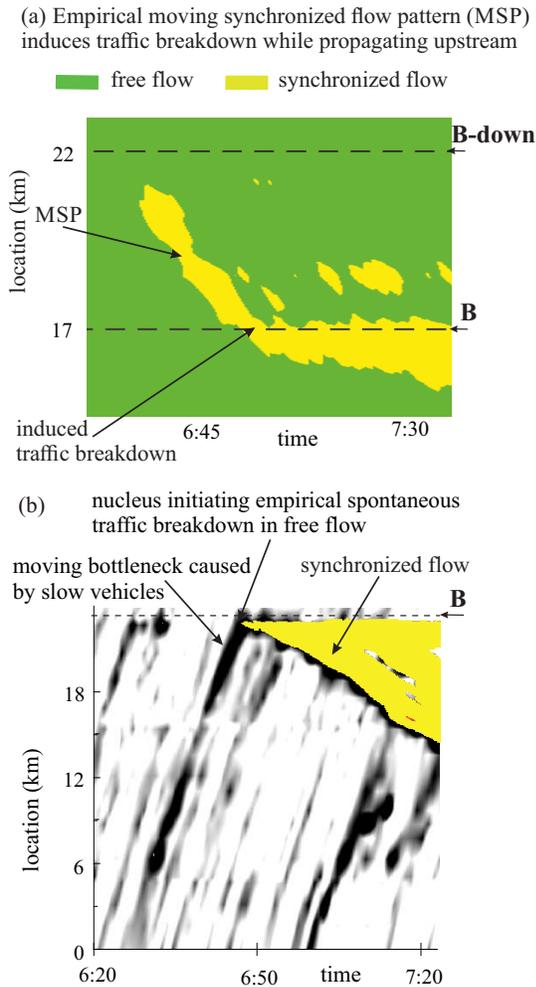}
\end{center}
\caption[]{Empirical nucleation nature of  traffic breakdown (F$\rightarrow$S transition) at bottlenecks in  human-driving vehicular traffic; traffic data were measured
with road detectors installed along   road sections~\cite{Kerner1999A,Kerner1999B,KernerBook,Kerner_Leibel}: 
(a) Speed data    
in space and time presented  with averaging method 
 of~\cite{KernerTomTom_2013}: A moving synchronized flow pattern (MSP) that has emerged   at   downstream bottleneck (B-down) while propagating upstream induces
F$\rightarrow$S transition (induced traffic breakdown) at upstream on-ramp bottleneck (B).
(b) One of the empirical waves (black colored waves) of
   decrease in the
average speed caused by slow moving vehicles (moving bottleneck) while propagating downstream in free flow 
acts as a nucleus for
spontaneous F$\rightarrow$S transition 
(spontaneous  traffic breakdown) at   bottleneck (B) when the speed wave propagates through   bottleneck B1.
Adapted from~\cite{KernerBook4}.
}
\label{Nucleation_Emp}
\end{figure}

However,	  from a study of empirical field traffic data   it was found that real  traffic breakdown is a transition from free flow (F) to synchronized flow (S) called an F$\rightarrow$S transition that occurs in metastable free flow with respect to an F$\rightarrow$S transition at a bottleneck~\cite{Kerner1999A,Kerner1999B,Kerner1998B}
(see for a review~\cite{KernerBook,KernerBook2,KernerBook3,KernerBook4,Kerner2019E}): The F$\rightarrow$S transition (traffic breakdown) exhibits the empirical nucleation nature
(Fig.~\ref{Nucleation_Emp}).   
	The LWR theory~\cite{LighthillWhitham,Richards,May,Daganzo1997,Manual2010,Roess2014,Ni2015}  cannot  explain
	 the   nucleation nature of real traffic breakdown. The classical traffic instability~\cite{Chandler,GH195910,Gazis1961A10,GH10,KS1,KS2,KS3,KS4,Newell1961,Newell1963A,Gipps,NSch1992,Bando1994,Reviews1,Reviews2,Reviews3,Reviews4,Saifuzzaman2015A,Schadschneider_Book,Treiber2013,Ashton1966,GerloughHuber1975,Gazis2002,Barcelo2010,Elefteriadou2014,Roess2014,Ni2015,Kessels2019,KernerKon1993} that leads to the F$\rightarrow$J transition~\cite{Reviews1,Reviews2,Reviews3,Reviews4,Treiber2013,KernerKon1993} cannot also explain real traffic breakdown at highway bottlenecks~\footnote{In more details, this criticism of standard traffic models can be found in books~\cite{KernerBook,KernerBook2,KernerBook3,KernerBook4}; 
	in particular, see Appendix~C in~\cite{KernerBook4}.}.

	To explain the empirical nucleation nature of traffic breakdown (F$\rightarrow$S transition), the author introduced  three-phase traffic theory~\cite{Kerner1999A,Kerner1999B,Kerner1998B,KernerBook,KernerBook2,KernerBook3,KernerBook4,Kerner2019E}). The three-phase traffic theory is a framework for the description of empirical traffic data in   three    phases: Free flow (F), synchronized flow (S) and wide moving jam (J); the traffic phases S and J belong to congested traffic.   The first implementations of the three-phase traffic theory in mathematical traffic flow models have been made in~\cite{KKl,KKW}. These
	stochastic   models have been further developed for different applications (see, e.g.,~\cite{KKl2003A}). Over time, other traffic flow models, which incorporate  hypotheses of the three-phases traffic theory,  have also been developed  (see, e.g.,~\cite{Davis,Lee_Sch2004A,Jiang2004A,Li2007,Hausken2015A,Tian_En2018,Junwei_Zeng2021A,Han-Tao_Zhao2020A,Wu2008,YangLu2013A,Yang2018A,Siebel2006,Rempe2017A,ReKl,Rehborn2011A,Rehborn_K_K_2020A,Qian2017A,Neto2011,Xiaojian_Hu2021B,Fu2020C,Kimathi2012B,Wiering2022A}). With the use of a microscopic three-phase traffic model for human-driving vehicles, the effect of 
	a small share of automated vehicles on traffic breakdown in mixed traffic at bottlenecks  has been studied in~\cite{KernerTPACC}.
	
\begin{figure}
\begin{center}
\includegraphics[width = 8 cm]{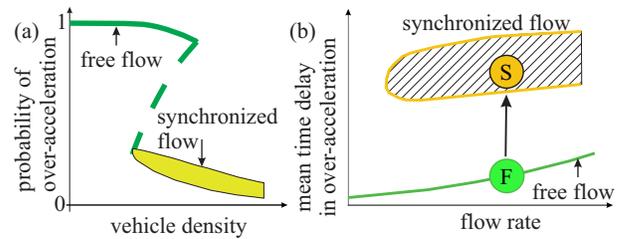}
\end{center}
\caption[]{Discontinuous character of
   over-acceleration~\cite{Kerner1999A,Kerner1999B,KernerBook}: 
(a) Qualitative presentation of  over-acceleration  probability during a given time interval. Equivalent 
	presentation of (a) as a discontinuous dependence of the mean time delay
in   over-acceleration on the flow rate; F and S are   states of free flow
and synchronized flow, respectively.
Adapted from~\cite{KernerBook,KernerBook4}.
}
\label{MeanTD_Over}
\end{figure}

	A basic hypothesis of the three-phase traffic theory is
that in some traffic situations vehicle acceleration  called {\it over-acceleration} exhibits  
a  {\it discontinuous  character}  (Fig.~\ref{MeanTD_Over}): In synchronized flow, the probability of over-acceleration
   is considerably lower than it is in free flow~\cite{Kerner1999A,Kerner1999B,KernerBook}~\footnote{See explanations of the term {\it over-acceleration} in Sec.~8.1.5 of the book~\cite{KernerBook4}.}.
It has been shown that the discontinuous character of over-acceleration 
causes a metastability of free flow with respect to
the F$\rightarrow$S transition; in its turn, this  metastability explains the empirical nucleation nature of traffic breakdown
observed in measured field traffic data. 
The three-phase traffic theory has been initially created 
	for the description of empirical {\it human-driving} vehicular traffic~\cite{Kerner1999A,Kerner1999B,Kerner1998B,KernerBook,KernerBook2,KernerBook3,KernerBook4}.  
	
	The objective of this paper is to show that the
	spatiotemporal dynamics
	of  traffic consisting of 100$\%$  automated-driving vehicles is described in the framework
	of three-phase traffic theory.
	It should be emphasized that  dynamic rules of motion of automated vehicles in a road lane can be developed that are
 totally different from the real dynamic behavior of human-driving vehicles.  Therefore, a question can arise:
 \begin{itemize}
\item Why should
 the three-phase traffic theory  describe spatiotemporal phase transitions in traffic flow consisting of 100$\%$ of automated   vehicles whose  
dynamics rules of motion in road lane can be totally different from the real dynamic behavior of human-driving vehicles?
  \end{itemize}

To answer   this question, we should recall that one of the mechanisms of over-acceleration  exhibiting the discontinuous character
(Fig.~\ref{MeanTD_Over}) is vehicle acceleration  through  lane-changing to a faster lane on a multi-lane road~\cite{Kerner1999A,Kerner1999B,KernerBook}~\footnote{In~\cite{Kerner1999A,Kerner1999B,KernerBook}, the probability of over-acceleration  shown in Fig.~\ref{MeanTD_Over} (a) 
has been called $\lq\lq$probability of passing" (see Fig.~5.7 (b) of the book~\cite{KernerBook}).}. Either
a human-driving or automated-driving vehicle changes to a neighborhood target lane if 
(i)   some {\it incentive conditions} for lane changing (like the vehicle can pass the preceding vehicle or/and move faster in the target lane) 
  {\it and} (ii) some {\it  safety conditions} 
	for lane-changing are satisfied, at which no  collisions
	between vehicles can occur.   Thus, if
		the discontinuous character of over-acceleration  due to lane-changing to a faster lane (Fig.~\ref{MeanTD_Over}) 
		is realized for human-driving vehicles, it should be also
		for automated vehicles: The discontinuous character of over-acceleration can be assumed to
   be   an universal physical feature  of vehicular traffic.

	The paper is organized as follows: In Sec.~\ref{Meta_S}, we consider a microscopic model of automated-driving vehicular traffic   on a two-lane road and study the physics of the nucleation nature of the F$\rightarrow$S
	transition at a bottleneck. The existence of a range of highway capacities at any time instant is the subject of Sec.~\ref{Range_S}.
In Sec.~\ref{Generalization_Sec}, a generalization of   nucleation features     of the  F$\rightarrow$S transition in 
	automated-driving traffic is made. Transitions between the three phases F, S, and J  in automated-driving traffic are studied in
	Sec.~\ref{WMJ_Sec}. In discussion (Sec.~\ref{Dis_Sec}), we show that
	the basic result about the nucleation nature of the F$\rightarrow$S transition at the bottleneck remains
	for string-unstable 
automated-driving traffic and even if a   different model for   automated-driving vehicles
is used.

  \section{Physics of metastability of automated-driving vehicular traffic  at bottleneck with respect to  F$\rightarrow$S
	transition 
	\label{Meta_S}}

\subsection{Model of automated-driving vehicular traffic on two-lane road with on-ramp bottleneck \label{Model_Sec}}

We study a    model of vehicular traffic consisting of 100$\%$
identical automated vehicles moving on a two-lane road with an
on-ramp bottleneck.
We assume that the control over an automated vehicle moving in a road lane   is realized through  
an adaptive cruise control system (ACC) that is described by a  classical  model in which the 
acceleration (deceleration) $a$ of
the automated vehicle is  determined  by   the space gap to the preceding vehicle $g=x_{\ell}-x-d$  
and the relative speed $\Delta v=v_{\ell}-v$ measured by the automated vehicle
 as well as by some optimal space gap $g_{\rm opt}$
  between the automated vehicle and the 
 preceding 
automated vehicle   (see, e.g.,~\cite{IoannouChien2002A,Levine1966A_Aut,Liang1999A_Aut,Liang2000A_Aut,Swaroop1996A_Aut,Rajamani2012A_Aut,Davis2004B9,Davis2014C}):
\begin{equation}
a = K_{1}(g-g_{\rm opt})+K_{2}\Delta v,  
 \label{ACC_General}
 \end{equation} 
where $x$ and $v$ are the coordinate and the speed of the automated vehicle, $x_{\ell}$ and
$v_{\ell}$ are the coordinate and the speed of the preceding automated vehicle,    $d$ is
	the vehicle length; here and below $v$, $v_{\ell}$, and $g$ are time-functions; $K_{1}$ and $K_{2}$ are  constant coefficients of  
   vehicle adaptation; 
	\begin{equation}
g_{\rm opt} =  v\tau_{\rm d},
 \label{ACC_g_opt}
 \end{equation} 
$\tau_{\rm d}$ is a desired time headway
of the automated vehicle to the 
 preceding 
automated vehicle. The classical model   (\ref{ACC_General}), (\ref{ACC_g_opt}) that is currently
	used in most studied of automated-driving in a road lane~\cite{IoannouChien2002A,Levine1966A_Aut,Liang1999A_Aut,Liang2000A_Aut,Swaroop1996A_Aut,Rajamani2012A_Aut,Davis2004B9,Davis2014C}
	is related to Helly's car-following model~\cite{Helly_1959}.
The motion of the automated vehicle in a road lane is found under conditions $0 \leq v \leq v_{\rm free}$
from the solution of   equations
$dv/dt=a$,  $dx/dt=v$~\footnote{These equations    are solved with the second-order Runge-Kutta method with time step 
$10^{-2}$. No noticeable changes in simulation results have been found when time step of calculation has been reduced to
$10^{-3}$ s.},
where the maximum speed (in free flow) $v_{\rm free}$ is a constant. 
There can be 
   string instability of a long enough platoon of   automated vehicles 
	(\ref{ACC_General}), (\ref{ACC_g_opt})~\cite{IoannouChien2002A,Levine1966A_Aut,Liang1999A_Aut,Liang2000A_Aut,Swaroop1996A_Aut,Rajamani2012A_Aut,Davis2004B9,Davis2014C}. As found by
	Liang and Peng~\cite{Liang1999A_Aut}, coefficients $K_{2}$ and $K_{1}$
	in (\ref{ACC_General}) can be chosen to  satisfy
condition for string stability 
	\begin{equation}
K_{2}>(2-K_{1}\tau_{\rm d}^{2})/2\tau_{\rm d}.
 \label{ACC_stability}
 \end{equation}   
 In the main text of the paper (Secs.~\ref{Meta_S}--\ref{WMJ_Sec}),  
we consider   only   automated vehicles whose parameters satisfy condition
(\ref{ACC_stability}) for string stability\footnote{The exclusion is   a discussion   in Sec.~\ref{String-unstable_Sec}, in which we compare    traffic phenomena at the bottleneck under string-stability condition (\ref{ACC_stability}) 
   studied in Secs.~\ref{Meta_S}--\ref{WMJ_Sec} with traffic phenomena occurring in free flow at the same bottleneck, when 
automated vehicles do not satisfy condition for string stability (\ref{ACC_stability}). }.

We use incentive lane changing rules from the right to left lane R$\rightarrow$L
(\ref{RL}) and from the left to right lane L$\rightarrow$R
(\ref{LR}) as well as safety conditions (\ref{g_prec_ACC}) known for human-driving vehicles (see, e.g.,~\cite{Nagel1998}) 
\begin{eqnarray}
\label{RL}
R \rightarrow L: v^{+}(t) \geq v_{\ell}(t)+\delta_{1} \ {\rm and} \ v(t) \geq v_{\ell}(t), \\
L \rightarrow R: v^{+}(t) \geq v_{\ell}(t)+\delta_{2} \ {\rm or} \  v^{+}(t) \geq v(t)+\delta_{2},
\label{LR} \\
g^{+}(t)   \geq\ v(t) \tau_{2}, \quad g^-(t) \geq\ v^{-}(t) \tau_{1},
\label{g_prec_ACC}
\end{eqnarray} 
at which
the automated vehicle changes to the faster target lane 
with the objective to pass a slower automated vehicle in the current lane     if   time headway to  preceding and following vehicles in the target  lane   are not shorter than some given safety time headway $\tau_{1}$
and $\tau_{2}$. In   (\ref{RL})--(\ref{g_prec_ACC}),
superscripts $+$  and $-$ denote, respectively, the preceding and the following vehicles in the target lane;
 $\tau_{1}$, $\tau_{2}$,   $\delta_{1}$, $\delta_{2}$  are positive constants.
  
Open boundary conditions are applied. At the beginning of the two-lane  road $x=0$ vehicles  are generated one after another
in each of the lanes of the road at time instants
$t^{(k)}=k\tau_{\rm in}$, $k=1,2,\ldots$, 
where $\tau_{\rm in}=1/q_{\rm in}$, $q_{\rm in}$ is a given time-independent flow rate   per road lane.
The initial vehicle speed is equal to  
$v_{\rm free}$. After a vehicle has reached the end of the road $x=L$ it is removed.
Before this occurs, the farthest downstream vehicle  maintains its speed and lane.

In the on-ramp model, there is
a merging region of length $L_{\rm m}$ in the right road lane that begins at road location $x=x_{\rm on}$
within which automated vehicles can merge from the on-ramp.  
  Vehicles
are generated at the on-ramp one after another at time instants
$t^{(m)}=m\tau_{\rm on}$, $m=1,2,\ldots$, 
where $\tau_{\rm on}=1/q_{\rm on}$, $q_{\rm on}$ is the on-ramp inflow rate.  To reduce
a  local  speed decrease occurring through the vehicle merging at the on-ramp bottleneck, 
as assumed for many known cooperative automated driving scenarios,     automated vehicles merge   with the speed of the preceding vehicle $v^{+}$ at a middle location
$x=(x^{+}+x^{-})/2$ between  the preceding and following vehicles in the right lane, when the space gap between the  vehicles 
  exceeds some safety value $g^{\rm (min)}_{\rm target}=\lambda_{\rm b}v^{+}+ d$, i.e., some safety condition
$x^{+}-x^{-}-d>g^{\rm (min)}_{\rm target}$ should be
  satisfied. In accordance with these merging conditions,
 the space gap for a vehicle merging between each pair of consecutive vehicles in the right road lane is checked, starting from the upstream boundary of the merging region. If there is such a pair of consecutive vehicles, the vehicle merges onto the right road lane;
 if there is no pair of consecutive vehicles, for  which the safety condition   is satisfied at the current time step, the procedure is repeated at the next time step,
and so on.

\begin{figure}
\begin{center}
\includegraphics[width = 8 cm]{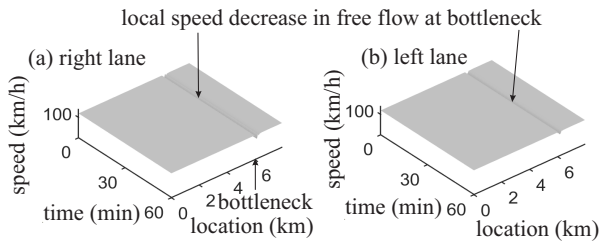}
\end{center}
\caption[]{Simulations  with model of Sec.\ref{Model_Sec} 
of  the occurrence of local speed decrease  in free flow on two-lane road
at    bottleneck: 
Speed in space and time in the right lane (a) and the left lane (b). 
 $q_{\rm in}=$ 2571 (vehicles/h)/lane, $q_{\rm on}=$ 720 vehicles/h. Parameters of automated vehicles: 
$\tau_{\rm d}=$ 1 s,
$K_{1}= 0.3 \ s^{-2}$, $K_{2}= 0.9 \ s^{-1}$, $v_{\rm free}=$ 120 km/h, $d=$ 7.5 m. Lane-changing parameters:
$\delta_{1}=$ 1 m/s, $\delta_{2}=$ 5 m/s, $\tau_{1}=$ 0.6 s, $\tau_{2}=$ 0.2 s. Road and on-ramp parameters: 
  road length $L=$ 8 km,   $x_{\rm on}=$ 6 km,  
$L_{\rm m}=$ 0.3 km, $\lambda_{\rm b}=$ 0.3 s.
}
\label{Free_Flow_Bottl} 
\end{figure} 

   When free flow is realized at the bottleneck,
we have found a known result that  due to R$\rightarrow$L lane-changing  
   the on-ramp inflow 
 is distributed between two lanes that
 causes the occurrence of
  local speed decreases in both the right and left road lanes at the bottleneck  (Fig.~\ref{Free_Flow_Bottl}).
		
\subsection{Free flow metastability  at bottleneck \label{Free_S_Induced}}

As mentioned, rules of vehicle motion of Sec.~\ref{Model_Sec} as well as the occurrence
of local speed decreases in both road lanes at the bottleneck in free flow
are known in vehicular traffic  theory. Nevertheless, 
  we have revealed   that  the free flow state  at the bottleneck  shown in Fig.~\ref{Free_Flow_Bottl} is
	in a metastable state with respect to an F$\rightarrow$S transition.

 \begin{figure}
\begin{center}
\includegraphics[width = 8 cm]{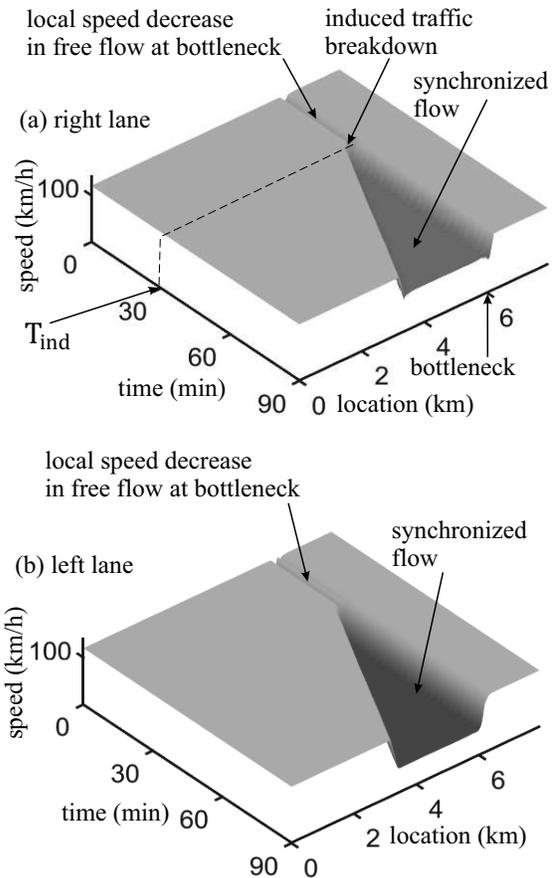}
\end{center}
\caption[]{Proof of the metastability of free flow state  shown in Fig.~\ref{Free_Flow_Bottl} in
automated-driving vehicular traffic moving on two-lane road with  bottleneck:
Speed in space and time in  the right lane (a) and left lane (b). Parameters of on-ramp inflow-rate impulse inducing F$\rightarrow$S transition   at bottleneck: $T_{\rm ind}=$ 30 min,
$\Delta q_{\rm on}=$ 180 vehicles/h, $\Delta t=$ 2 min.
 Other model parameters are the same as those in Fig.~\ref{Free_Flow_Bottl}.  
}
\label{Induced_F_S_Bottl}
\end{figure}

	To prove this   result, at a time instant $T_{\rm ind}$
	we have disturbed the free flow state at the bottleneck  shown in Fig.~\ref{Free_Flow_Bottl} through the application of
	a time-limited      on-ramp inflow   impulse $\Delta q_{\rm on}$  of  some duration $\Delta t$ 
(Fig.~\ref{Induced_F_S_Bottl}): (i) At time interval $0\leq t < T_{\rm ind}$, the   on-ramp inflow rate  $q_{\rm on}$ is
 the same as that in Fig.~\ref{Free_Flow_Bottl} and, therefore, the same free flow state
is realized; (ii) during the impulse
$T_{\rm ind} \leq t \leq T_{\rm ind} + \Delta t$ the on-ramp inflow rate has increased to a large enough value
  $q_{\rm on}+\Delta q_{\rm on}$ at which traffic congestion
  is realized at the bottleneck;
(iii) at time $t > T_{\rm ind} + \Delta t$, although the   on-ramp inflow rate has reduced to its initial value $q_{\rm on}$,
  rather the free flow state returns at the bottleneck,  congested traffic  persists  at the bottleneck.
	 The downstream front of induced congested traffic 
	is fixed at the bottleneck while
	the upstream   front of congested traffic is continuously propagate upstream (Fig.~\ref{Induced_F_S_Bottl}).
	 In accordance with the phase definitions  made in three-phase traffic theory~\cite{KernerBook}, the induced
	congested traffic  belongs to the synchronized flow phase of automated-driving vehicular traffic.   
 Thus, at the same on-ramp inflow rate   $q_{\rm on}$ there can be either a free flow state or a synchronized flow state
 at the bottleneck, i.e., free flow in Fig.~\ref{Free_Flow_Bottl}
	is indeed in a metastable state with respect to an F$\rightarrow$S transition at the bottleneck.

 \subsection{Discontinuity in the rate of over-acceleration through lane-changing   \label{Physics_meta_Sec}}

	To explain the physics of the free flow metastability   with respect to the F$\rightarrow$S transition
  (Sec.~\ref{Adaptation_Over_Sec}), we should first explain here that there is a {\it discontinuity} in the rate of R$\rightarrow$L lane-changing
denoted by $R_{\rm RL}$~\footnote{$R_{\rm RL}$ is the number of automated vehicles that change from the right lane to the left lane during a time unit within the road region $x_{\rm on}- L_{\rm RL}  \leq x \leq x_{\rm on}+L_{\rm m}$, where   parameter
 $L_{\rm RL}=$ 0.06--0.1 km used in simulations
 should guarantee that R$\rightarrow$L lane-changing at the upstream front of synchronized flow (after the F$\rightarrow$S transition has occurred, see Sec.~\ref{Effect_Dis_Sec}) does not
come in the calculation of $R_{\rm RL}$.}.
The discontinuity in the rate of R$\rightarrow$L lane-changing is realized due to the F$\rightarrow$S transition, i.e.,
when free flow transforms into synchronized flow.
Examples of  R$\rightarrow$L lane-changing in free flow and synchronized flow are shown, respectively,  
in Figs.~\ref{Free_Flow_Bottl_tr} and~\ref{Induced_F_S_Bottl_tr}
 through the use of dashed vertical lines R$\rightarrow$L. In free flow
occurring during time $0\leq t < T_{\rm ind}$, we have found
$R_{\rm RL}\approx$ 6.1 $\rm min^{-1}$, whereas in synchronized flow
that occurs at the bottleneck at $t > T_{\rm ind} + \Delta t$, we have found that  
 R$\rightarrow$L lane-changing rate $R_{\rm RL}$ reduces sharply to
  $R_{\rm RL}\approx$ 2.8 $\rm min^{-1}$. To explain the abrupt reduction of the rate of 
	R$\rightarrow$L lane-changing   $R_{\rm RL}$ occurring due to the F$\rightarrow$S transition, we  mention that
in synchronized flow, the mean time headway between vehicles $\tau^{\rm (syn)}_{\rm mean}$ reduces
 while becoming close to   $\tau_{\rm d}=$ 1 s. At this short time headway, 
safety conditions for lane changing (\ref{g_prec_ACC}) is more difficult
to satisfy in comparison with   free flow for which $\tau^{\rm (free)}_{\rm mean}\approx$ 1.175 s~\footnote{The mean time headway between vehicles
in free flow is equal to $\tau^{\rm (free)}_{\rm mean}=(3600/q_{\rm in})- (d/v_{\rm free})\approx$ 1.175 s.  }.
	The difference in values of $R_{\rm RL}$ in free flow and synchronized flow can already be seen from a comparison of two
	 fragments of vehicle trajectories in the vicinity of the on-ramp
	merging region shown for free flow in 
	Fig.~\ref{Free_Flow_Bottl_tr}~\footnote{It should be emphasized 
	that   
the free flow state at the bottleneck shown in Fig.~\ref{Induced_F_S_Bottl}
during   time interval $0\leq t < T_{\rm ind}$  is {\it identical} with the free flow state at the bottleneck
in Fig.~\ref{Free_Flow_Bottl}. Therefore,
parameters of trajectories   shown in Fig.~\ref{Free_Flow_Bottl_tr} are related to both Fig.~\ref{Free_Flow_Bottl} and  
 to Fig.~\ref{Induced_F_S_Bottl} for   time interval $0\leq t < T_{\rm ind}$.} and for synchronized flow 
in Fig.~\ref{Induced_F_S_Bottl_tr}.

	\begin{figure}
\begin{center}
\includegraphics[width = 8 cm]{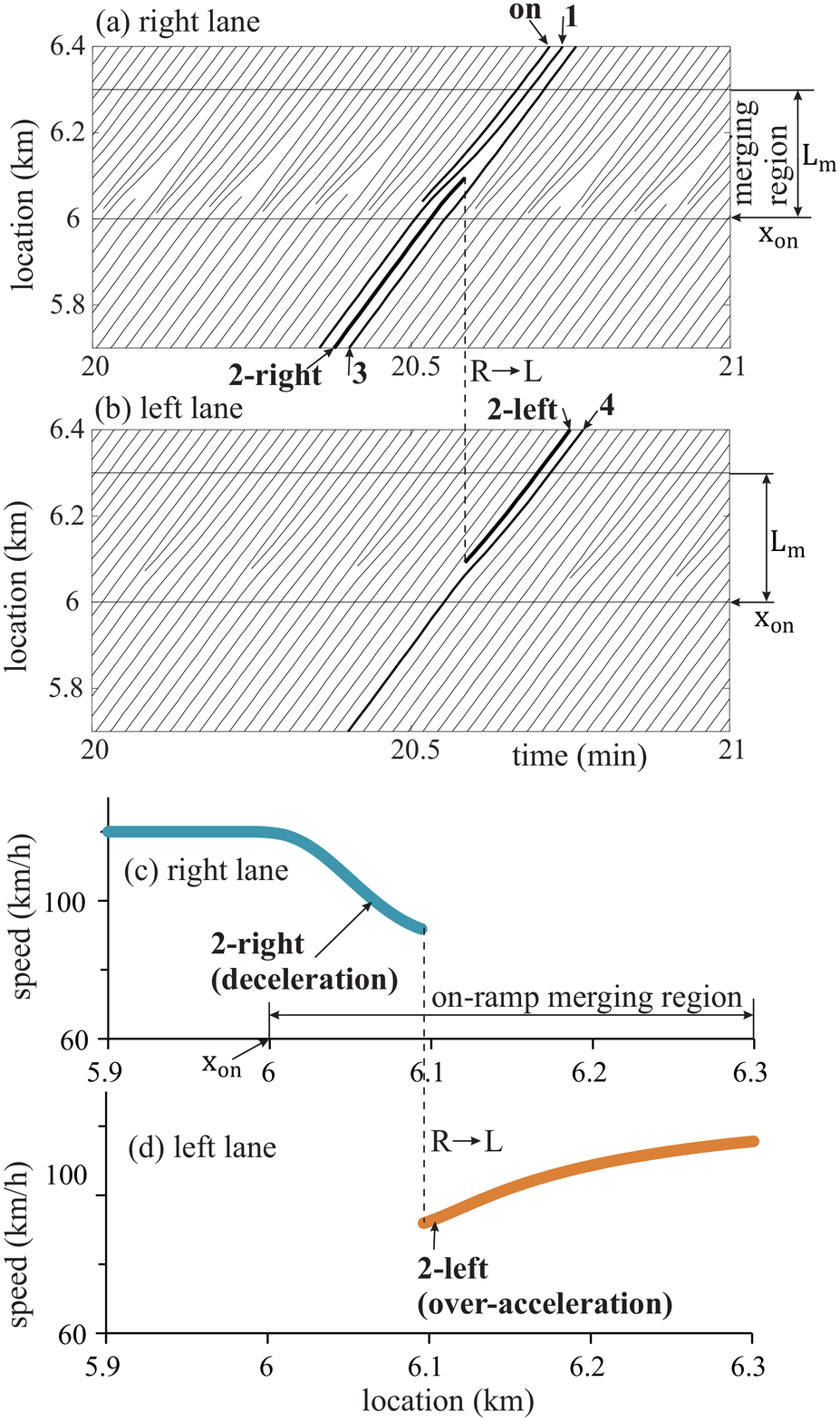}
\end{center}
\caption[]{Continuation of Figs.~\ref{Free_Flow_Bottl} and~\ref{Induced_F_S_Bottl}.  (a, b) Simulated vehicle trajectories within  local speed decrease in free flow
at  bottleneck in  the right lane (a) and left lane (b) at time $t < T_{\rm ind} + \Delta t$.
 (c, d)
Location-functions of speed  of vehicle 2 labeled by $\lq\lq$2-right" in the right lane (c) and by
$\lq\lq$2-left" in left lane 
 (d)      in (a, b).
R$\rightarrow$L lane-changing of vehicle 2  is marked by   dashed vertical lines R$\rightarrow$L.    
}
\label{Free_Flow_Bottl_tr}
\end{figure}

	\begin{figure} 
\begin{center}
\includegraphics[width = 8 cm]{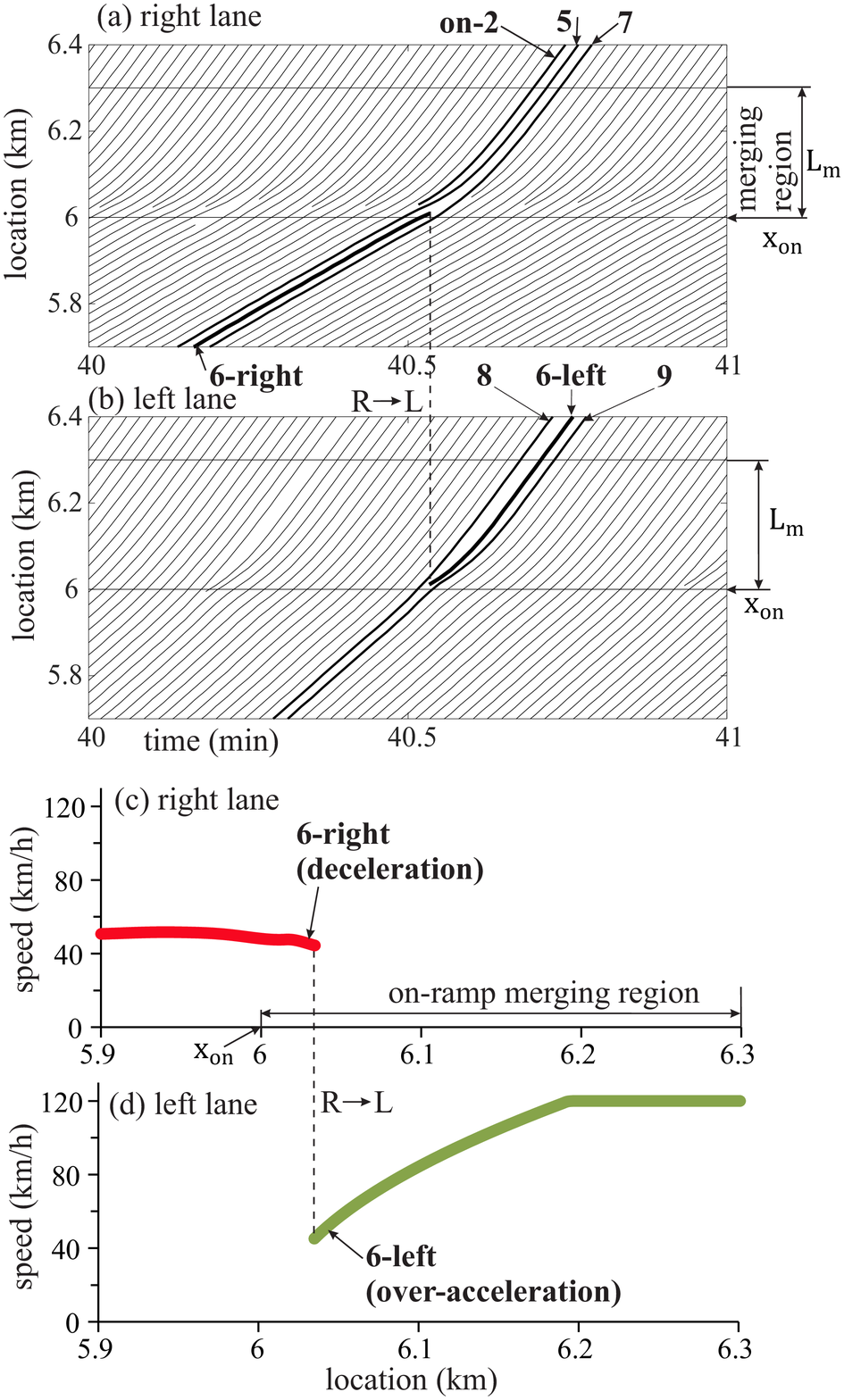}
\end{center}
\caption[]{Continuation of Fig.~\ref{Induced_F_S_Bottl}.    (a, b) Simulated vehicle trajectories in synchronized flow
at  bottleneck in the right lane (a) and left lane (b) at time $t > T_{\rm ind} + \Delta t$.
 (c, d)
Location-functions of speed  of vehicle 6 labeled by $\lq\lq$6-right" in the right lane (c) and by
$\lq\lq$6-left" in left lane 
 (d)      in (a, b).
R$\rightarrow$L lane-changing of vehicle 6  is marked by   dashed vertical lines R$\rightarrow$L.   
}
\label{Induced_F_S_Bottl_tr}
\end{figure}

  R$\rightarrow$L lane-changing of a vehicle that has initially decelerated in the right lane
(for example, vehicle 2-right in Figs.~\ref{Free_Flow_Bottl_tr} (a, c) and vehicle
 6-right in Figs.~\ref{Induced_F_S_Bottl_tr} (a, c) have decelerated before R$\rightarrow$L lane-changing) leads to
the acceleration of the vehicle in the left lane. Indeed,
in free flow, vehicle 2-left in Figs.~\ref{Free_Flow_Bottl_tr} (b, d)   accelerates
 after R$\rightarrow$L lane-changing. In synchronized flow, vehicle 6-left in Figs.~\ref{Induced_F_S_Bottl_tr} (b, d)
also	accelerates
 after R$\rightarrow$L lane-changing. The vehicle acceleration under consideration is solely determined
		by    R$\rightarrow$L lane-changing of the vehicle. Thus, 
		the rate of   the vehicle acceleration  denoted by $R_{\rm OA}$, which is  caused by R$\rightarrow$L lane-changing, is given by
		formula
		\begin{equation}
	R_{\rm OA}=R_{\rm RL}. 
	\label{R_OA_R_F}
	\end{equation}
		  Thus,	vehicle acceleration caused by R$\rightarrow$L lane-changing exhibits the discontinuous character: In 
	  accordance with (\ref{R_OA_R_F}), there is the
	discontinuity in the rate
	of   vehicle acceleration   $R_{\rm OA}$   when free flow transforms into synchronized flow.

	In next Sec.~\ref{Adaptation_Over_Sec}, we explain that
	the discontinuity in   the rate of
	  vehicle acceleration $R_{\rm OA}$ caused by R$\rightarrow$L lane-changing
	 leads to the
			free flow metastability   with respect to the F$\rightarrow$S transition
  (Fig.~\ref{Induced_F_S_Bottl});
	in   three-phase traffic theory, such vehicle acceleration  
	  has been called {\it over-acceleration}~\cite{KernerBook,KernerBook4}.
	Therefore, the acceleration of  vehicle 2-left in free flow in Figs.~\ref{Free_Flow_Bottl_tr} (b, d)
		as well as the acceleration of  vehicle 6-left in synchronized flow
	  (Figs.~\ref{Induced_F_S_Bottl_tr} (b, d)) are examples of over-acceleration;
		this explains the use of the term {\it over-acceleration} in Figs.~\ref{Free_Flow_Bottl_tr} and~\ref{Induced_F_S_Bottl_tr}.
			We consider also the mean time delay in over-acceleration denoted by 
	$T_{\rm OA}$ that is equal to $1/R_{\rm OA}$;  in free flow
	$T_{\rm OA}\approx$ 9.84 s, whereas in synchronized flow $T_{\rm OA}\approx$ 21.4 s.
	The discontinuities in the rate $R_{\rm OA}$ and mean time delay $T_{\rm OA}$ of over-acceleration,
		i.e., the discontinuous character of over-acceleration
	found here for automated-driving vehicular traffic
	is in   agreement with   three-phase traffic theory for human-driving traffic
 (Fig.~\ref{MeanTD_Over}).

\subsection{Spatiotemporal competition of speed adaptation with  over-acceleration    \label{Adaptation_Over_Sec}}

	There is a spatiotemporal competition between over-acceleration  
	and speed adaptation.
	In this competition, there are a tendency to free flow and the opposite
	tendency to synchronized flow. The tendency to free flow is through over-acceleration.
	The opposite tendency to synchronized flow  is through speed adaptation.

Speed adaptation is vehicle deceleration occurring when a   vehicle approaches
a slower moving preceding  vehicle and the following vehicle cannot pass it.
We should distinguish speed adaptation  in the right lane and
speed adaptation in the left lane. This is because speed adaptation in the left lane
is caused by a dual role of R$\rightarrow$L lane-changing.

\subsubsection{Tendency to free flow   through over-acceleration}

In free flow (Fig.~\ref{Free_Flow_Bottl_tr}) and   synchronized flow
(Fig.~\ref{Induced_F_S_Bottl_tr}),  over-acceleration through vehicle lane-changing to the left lane
permits the  following vehicle remaining in the right lane to accelerate.   
When  free flow is currently at the bottleneck, the  tendency to free flow is through over-acceleration
that maintains
the free flow state. Indeed, due to over-acceleration of vehicle 2 through its changing to the left lane
 ($\lq\lq$2-left (over-acceleration)" in Fig.~\ref{Over_Following_tr} (a)), the following vehicle 3 remaining in the right lane
that trajectory is shown in Fig.~\ref{Free_Flow_Bottl_tr} (a) 
accelerates (labeled by $\lq\lq$3, acceleration"  in Fig.~\ref{Over_Following_tr} (a)).
When  synchronized flow is currently at the bottleneck, the tendency caused by over-acceleration tries to
transform synchronized flow to a
  free flow state. For example, due to over-acceleration of vehicle 6 through its changing to the left lane
 ($\lq\lq$6-left (over-acceleration)" in Fig.~\ref{Over_Following_tr} (b)) the following vehicle 7 remaining in the right lane
that trajectory is shown in Fig.~\ref{Induced_F_S_Bottl_tr} (a) 
accelerates (labeled by $\lq\lq$7, acceleration" in Fig.~\ref{Over_Following_tr} (b)).
 
	\begin{figure} 
\begin{center}
\includegraphics[width = 8 cm]{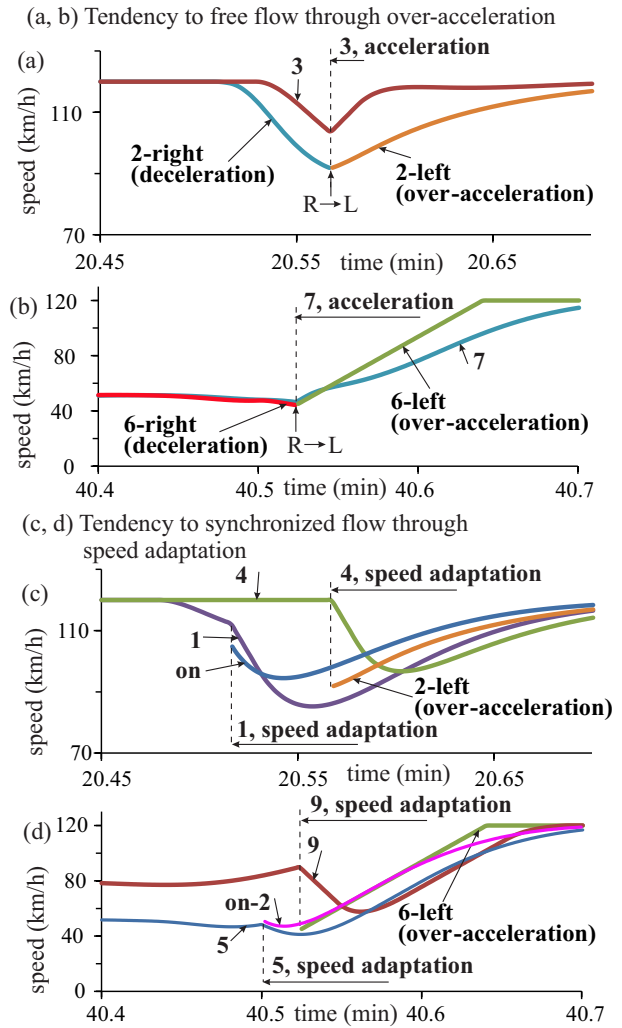}
\end{center}
\caption[]{Simulations of spatiotemporal competition between over-acceleration and speed adaptation.
Time-functions of speed for vehicle trajectories presented in Figs.~\ref{Free_Flow_Bottl_tr} (a, b) and~\ref{Induced_F_S_Bottl_tr} (a, b) labeled by the same numbers, respectively: (a, b) Tendency to free flow. (c, d) Tendency to synchronized flow.  
}
\label{Over_Following_tr}
\end{figure}

 \subsubsection{Tendency to synchronized flow    through speed adaptation in the right lane}

 When  free flow is  at the bottleneck, the tendency caused by speed adaptation tries to
transform free flow to a
  synchronized flow (Fig.~\ref{Over_Following_tr} (c)). A vehicle merging from the on-ramp 
(vehicle $\lq\lq$on" in Figs.~\ref{Free_Flow_Bottl_tr} (a) and~\ref{Over_Following_tr} (c)),
forces the following vehicle 1 that trajectory is shown in Fig.~\ref{Free_Flow_Bottl_tr} (a) 
  to decelerate
	($\lq\lq$1, speed adaptation" in Fig.~\ref{Over_Following_tr} (c)) 
	while adapting the speed to the slower merging vehicle   $\lq\lq$on".

	If  synchronized flow is   at the bottleneck, the tendency caused by speed adaptation tries to
maintain  the synchronized flow  state (Fig.~\ref{Over_Following_tr} (d)). A vehicle merging from the on-ramp 
(vehicle $\lq\lq$on-2" in Figs.~\ref{Induced_F_S_Bottl_tr} (a) and~\ref{Over_Following_tr} (d)),
forces the following vehicle 5 that trajectory is shown in Fig.~\ref{Induced_F_S_Bottl_tr} (a) 
  to decelerate
	(labeled by $\lq\lq$5, speed adaptation" in Fig.~\ref{Over_Following_tr} (d)).
 
 \subsubsection{Tendency to synchronized flow   through speed adaptation in the left lane:
Dual role of lane-changing}

 There is a dual role of lane-changing that is as follows.
In free flow,   
  lane-changing of vehicle 2
leads to over-acceleration   ($\lq\lq$2-left (over-acceleration)" in Figs.~\ref{Free_Flow_Bottl_tr} (a, d)).
Contrarily, the same
lane-changing of vehicle 2   causes speed adaptation in the left lane. Indeed,  
the following vehicle 4 in the left lane that trajectory is shown in Fig.~\ref{Induced_F_S_Bottl_tr} (b) 
  must decelerate
	($\lq\lq$4, speed adaptation" in Fig.~\ref{Over_Following_tr} (d)),
		while adapting its speed to the speed of   slower   vehicle 2    that has just changed from the right lane
	to   left  lane.

Speed adaptation caused by a dual role of lane-changing occurs also in synchronized flow.
An example is  
lane-changing of vehicle 6
 ($\lq\lq$6-left (over-acceleration)" in Fig.~\ref{Over_Following_tr} (d)) that forces
the following vehicle 9 in the left lane that trajectory is shown in Fig.~\ref{Induced_F_S_Bottl_tr} (b) 
  to decelerate
	($\lq\lq$9, speed adaptation" in Fig.~\ref{Over_Following_tr} (d)).

 \subsubsection{Two possible results of   competition  between over-acceleration and speed adaptation}
	
	In Fig.~\ref{Free_Flow_Bottl_tr}, free flow persists
	at the bottleneck. This means that at the   over-acceleration rate  $R_{\rm OA} \approx$ 6.1 $\rm min^{-1}$ 
	the  tendency to free flow through over-acceleration overcomes
	the tendency to synchronized flow through speed adaptation. 
 The result of the competition  between over-acceleration and speed adaptation
	is the occurrence of the local speed decrease at the bottleneck   without
	the emergence of synchronized flow  (Figs.~\ref{Free_Flow_Bottl} and~\ref{Induced_F_S_Bottl}).

	Contrarily, 	   in Fig.~\ref{Induced_F_S_Bottl_tr}
	synchronized flow persists
	at the bottleneck (labeled
	by $\lq\lq$synchronized flow").
	This means that   the tendency to synchronized flow through speed adaptation
	overcomes the tendency to free flow through over-acceleration. This is
	because   the   over-acceleration rate  
 $R_{\rm OA} \approx$ 2.8 $\rm min^{-1}$ becomes too small in synchronized flow.
 The competition between speed adaptation and over-acceleration determines
	the speed in  synchronized flow. However, due to the small rate of over-acceleration in synchronized flow
	this competition cannot cause a return transition from synchronized flow to free flow.
	
	Thus, the cause of  the free flow metastability   with respect to the F$\rightarrow$S transition
  (Fig.~\ref{Induced_F_S_Bottl})
  is a spatiotemporal competition between over-acceleration, which exhibits the discontinuous character,
	and speed adaptation.

\subsection{Synchronized flow characteristics
\label{Effect_Dis_Sec}}

 \subsubsection{Synchronization of velocities of upstream fronts of synchronized flow in road lanes  \label{Synch_Front_Sec}}

The speed  in synchronized flow in the right lane (vehicle 10 in Fig.~\ref{Induced_F_S_Bottl_tr_S}) is less that the speed in synchronized flow in the left lane
(vehicle 11). However, this speed  difference  does not lead to   different
velocities of the upstream fronts of synchronized flow in the left and right lanes: These upstream front velocities
are synchronized (upstream fronts of synchronized flow 
are labeled by dashed curves $\lq\lq$S-up" in Fig.~\ref{Induced_F_S_Bottl_tr_S} (a, b)). 

		\begin{figure} 
\begin{center}
\includegraphics[width = 8 cm]{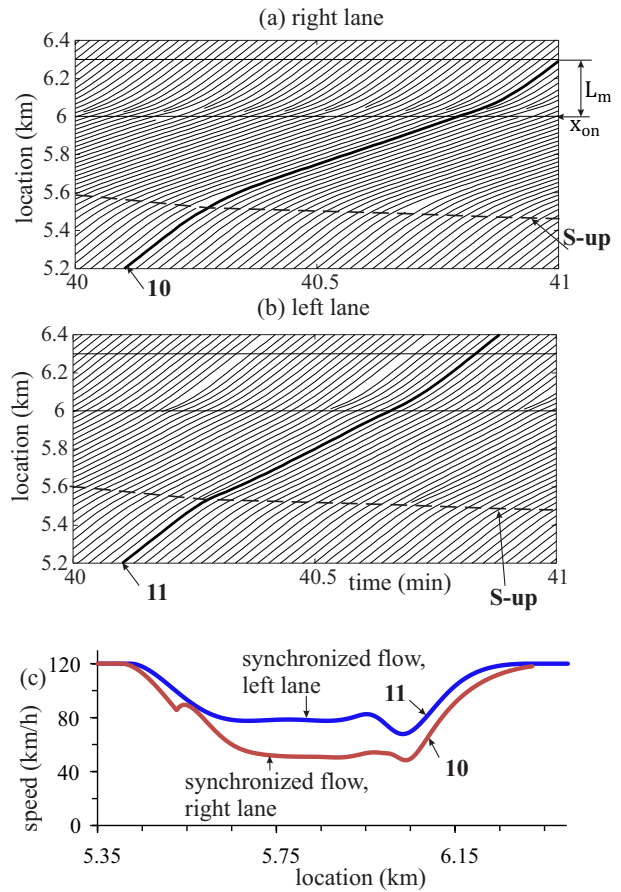}
\end{center}
\caption[]{Continuation of Fig.~\ref{Induced_F_S_Bottl_tr}.
Features of synchronized flow:
(a, b)   Vehicle trajectories at       $t > T_{\rm ind} + \Delta t$, i.e., after F$\rightarrow$S transition has occurred  at the bottleneck.
(c, d) Location-functions of  speeds  
 for vehicles 10 and 11 in (a, b). 
}
\label{Induced_F_S_Bottl_tr_S}
\end{figure}

	\begin{figure} 
\begin{center}
\includegraphics[width = 8 cm]{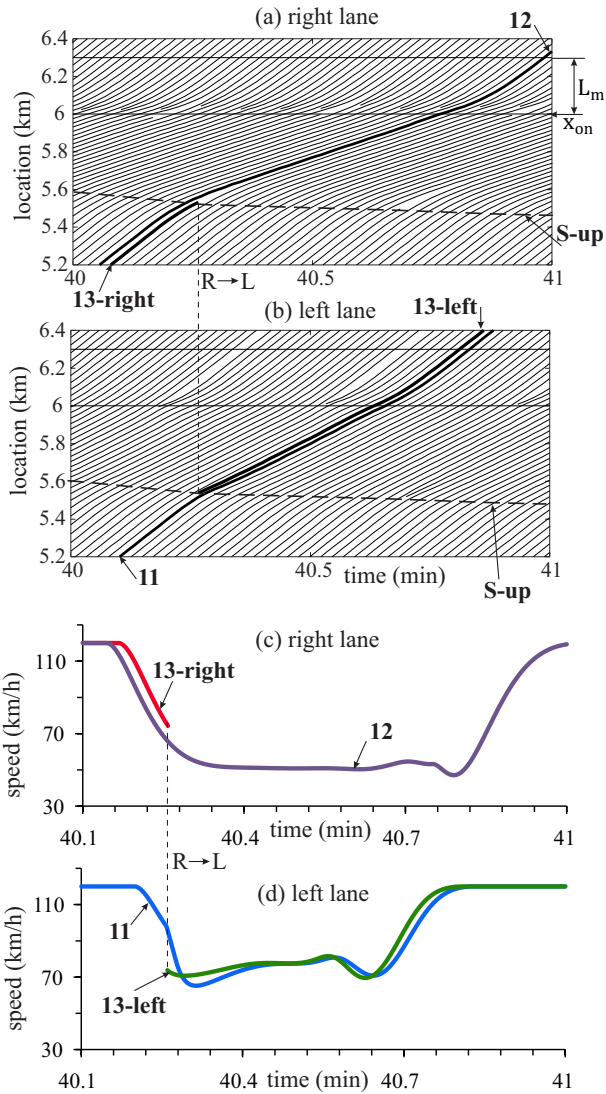}
\end{center}
\caption[]{Continuation of Fig.~\ref{Induced_F_S_Bottl_tr_S}:
Synchronization of velocities of upstream fronts of synchronized flow 
in the right and left lanes.  (a, b) Vehicle trajectories taken from Fig.~\ref{Induced_F_S_Bottl_tr_S} (a, b).
(c, d) Time-functions of  speeds along      trajectories marked in (a, b) by the same numbers, respectively.
}
\label{Induced_F_S_Bottl_tr_up}
\end{figure}

The physics of this synchronization effect  is associated with R$\rightarrow$L
 lane-changing 
 that occur in the vicinity of the upstream synchronized flow front in the right lane (Fig.~\ref{Induced_F_S_Bottl_tr_up}).   
While approaching the upstream front of synchronized flow in the right lane, vehicles decelerate (e.g., vehicle 12 in
Fig.~\ref{Induced_F_S_Bottl_tr_up} (a, c)). When the upstream front of synchronized flow in the left lane   comes even slightly  
downstream of the upstream front of synchronized flow in the right lane, free flow
is realized in the left lane 
between these upstream synchronized flow fronts; then, between the fronts lane-changing rate $R_{\rm RL}$  increases.
This causes  R$\rightarrow$L
  lane-changing of a vehicle decelerating to a synchronized flow speed
	in the vicinity of the upstream front of synchronized flow in the right lane (example of R$\rightarrow$L lane-changing for  
  vehicle 13 is marked by dashed vertical lines labeled by R$\rightarrow$L in Fig.~\ref{Induced_F_S_Bottl_tr_up}).
Due to the lane-changing of a slow moving vehicle 13-right to the left lane (vehicle 13-left), the following vehicle
11 in the left lane   
begins to decelerate stronger than it has been before lane-changing (Fig.~\ref{Induced_F_S_Bottl_tr_up} (d)).
This leads to the synchronization of the upstream front velocities.

\subsubsection{Effect of discontinuity in lane-changing rate on   flow-rate distribution  \label{Dis_flow_Sec}}

	\begin{figure} 
\begin{center}
\includegraphics[width = 8 cm]{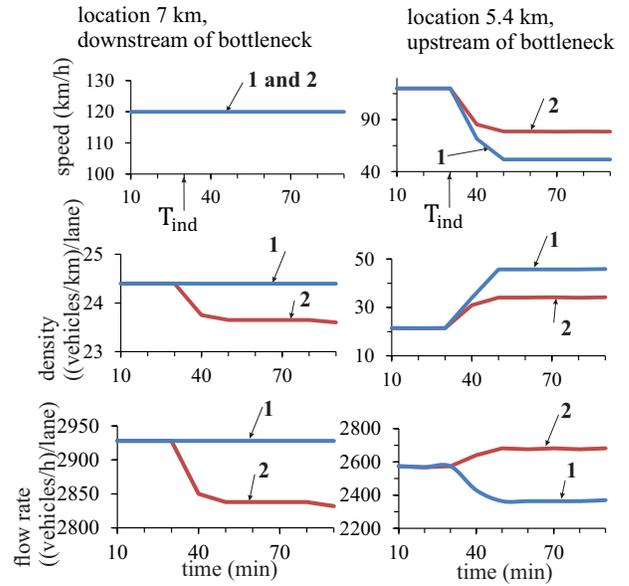}
\end{center}
\caption[]{Continuation of Fig.~\ref{Induced_F_S_Bottl}:
Time-functions of automated vehicle speed (first line),  density (second line), and flow rate (third line)
at road location $x=$ 7 km [downstream of the bottleneck] (left column)
and road location $x=$ 5.4 km [upstream of the bottleneck] (right column); curves 1 -- right lane, curves 2 -- left lane.
10 min averaging time interval at virtual detectors.
}
\label{Speed-Time-lane-symmetric}
\end{figure}

In the initial free flow state existing at the bottleneck at $0 \leq t < T_{\rm ind}$  
(Fig.~\ref{Induced_F_S_Bottl}),  R$\rightarrow$L lane-changing leads to the nearly fully equalization of the flow rates and densities between the road lanes downstream of the bottleneck (left column in
Fig.~\ref{Speed-Time-lane-symmetric} at $t< T_{\rm ind}=$ 30 min).  
After   the F$\rightarrow$S transition has occurred, 
 the lane-changing rate in synchronized flow at the bottleneck
 decreases sharply  (discontinuity in the lane-changing rate) and, therefore,
  the flow rates and densities between  lanes cannot be equalized. This explains why
 in free flow downstream   of the   bottleneck
both the density and flow rate are smaller in the left lane than they are, respectively, in the right lane
(left column in Fig.~\ref{Speed-Time-lane-symmetric}  at $t\geq T_{\rm ind}$).

 The discontinuity in the lane-changing rate is
also responsible for differences in the averaged
 speeds, densities, and flow rates in synchronized flow in the right and left lanes upstream of the bottleneck
(right column in Fig.~\ref{Speed-Time-lane-symmetric}   at $t\geq  T_{\rm ind}$).

\subsection{Three-phase traffic theory as   common framework for    human-driving and automated-driving
 traffic \label{Common-Sec}}

\begin{figure}
\begin{center}
\includegraphics[width = 8 cm]{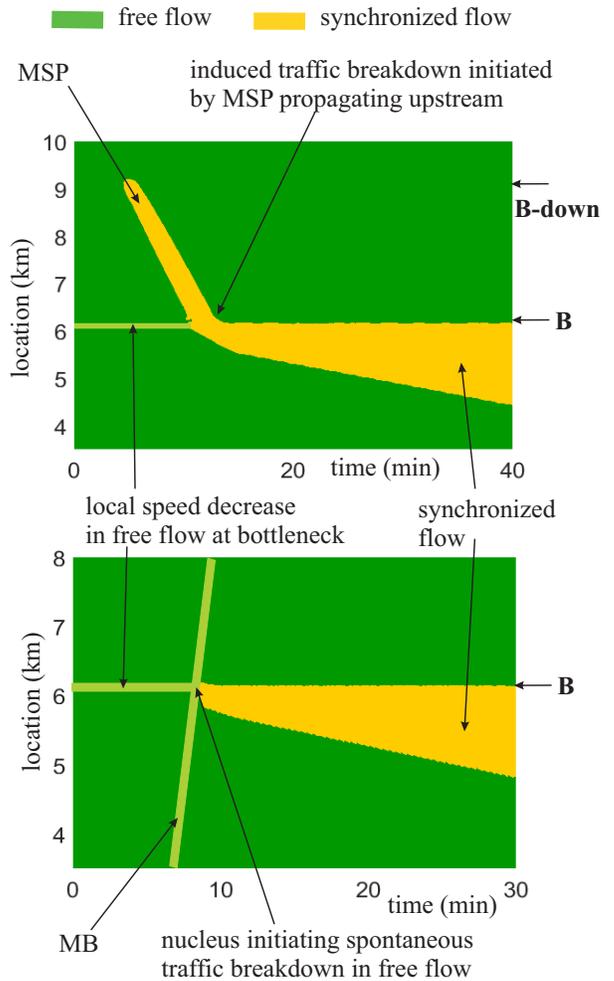}
\end{center}
\caption[]{Simulations of automated-driving vehicular traffic that reproduce empirical
   breakdown  nucleation features measured  in  real human-driving   traffic (Fig.~\ref{Nucleation_Emp}). 
  Speed data averaged across  two-lane road are
presented  in space and time in free flow (green) and synchronized flow (yellow): (a)
A moving synchronized flow pattern (MSP)  induced
at the downstream bottleneck (B-down)     propagates upstream; reaching
the upstream on-ramp bottleneck (B) the MSP induces the
F$\rightarrow$S transition at the bottleneck.
(b) A slow moving vehicle (moving bottleneck -- MB) while propagating downstream in free flow 
acts as a nucleus for
empirical spontaneous F$\rightarrow$S transition at   bottleneck B when the MB propagates through   bottleneck B.
Both bottleneck B-down and bottleneck B are identical with the   on-ramp bottleneck used 
 above  (Figs.~\ref{Free_Flow_Bottl}--\ref{Speed-Time-lane-symmetric});
more details of simulations are in Figs.~\ref{MSP_Breakdown} and~\ref{MB_Breakdown}. 
}
\label{MSP_MB_color}
\end{figure}

Simulations of  
 automated-driving vehicular traffic (Figs.~\ref{MSP_MB_color} (a, b))   show
the empirical
 nucleation features of the F$\rightarrow$S transition found in measurements of
 real human-driving traffic  (Figs.~\ref{Nucleation_Emp} (a, b)).
Thus, three-phase traffic theory can indeed be considered
a common framework for the analysis of the dynamics of  
  human-driving and automated-driving
 traffic.

\begin{figure}
\begin{center}
\includegraphics[width = 8 cm]{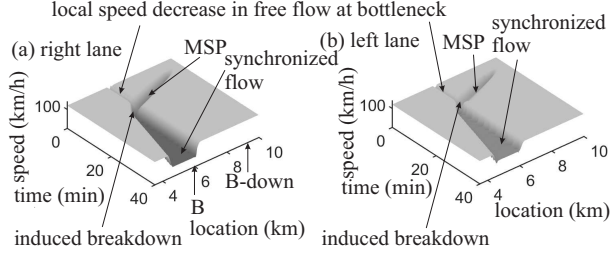}
\end{center}
\caption[]{Simulations of F$\rightarrow$S transition  through upstream propagation of MSP 
to upstream bottleneck  that in simplified version is shown in Fig.~\ref{MSP_MB_color} (a): Speed in space and time in the right
lane  (a) and  left lane
 (b). $q_{\rm in}=$ 2571
(vehicles/h)/lane. Two-lane road with two bottlenecks: Parameters of upstream bottleneck (B) are $x_{\rm on}=$ 
  6 km, $L_{\rm m}=$ 0.3 km, $q_{\rm on}=$ 720 vehicles/h; 
parameters of downstream bottleneck (B-down) are $x^{\rm (down)}_{\rm on}=$ 9 km,
$L^{\rm (down)}_{\rm m}=$ 0.3 km, $q^{\rm (down)}_{\rm on}=$ 0;
road length $L=$ 10 km.
Parameters of on-ramp inflow impulse  at downstream  bottleneck B-down applied at  $T^{\rm (down)}_{\rm ind}=$ 5 min are
 $\Delta q^{\rm (down)}_{\rm on}=$ 900 vehicles/h,
 $\Delta t^{\rm (down)}=$ 1 min.  
Other model parameters are the same as those in Fig.~\ref{Free_Flow_Bottl}. 
}
\label{MSP_Breakdown}
\end{figure}

A moving synchronized flow pattern (MSP) in Fig.~\ref{MSP_MB_color} (a)  has been induced through  the  use of an on-ramp inflow impulse at a downstream  bottleneck   (B-down in Fig.~\ref{MSP_Breakdown}).
While propagating upstream, the MSP induces the
F$\rightarrow$S transition at the upstream   bottleneck.

 \begin{figure}
\begin{center}
\includegraphics[width = 8 cm]{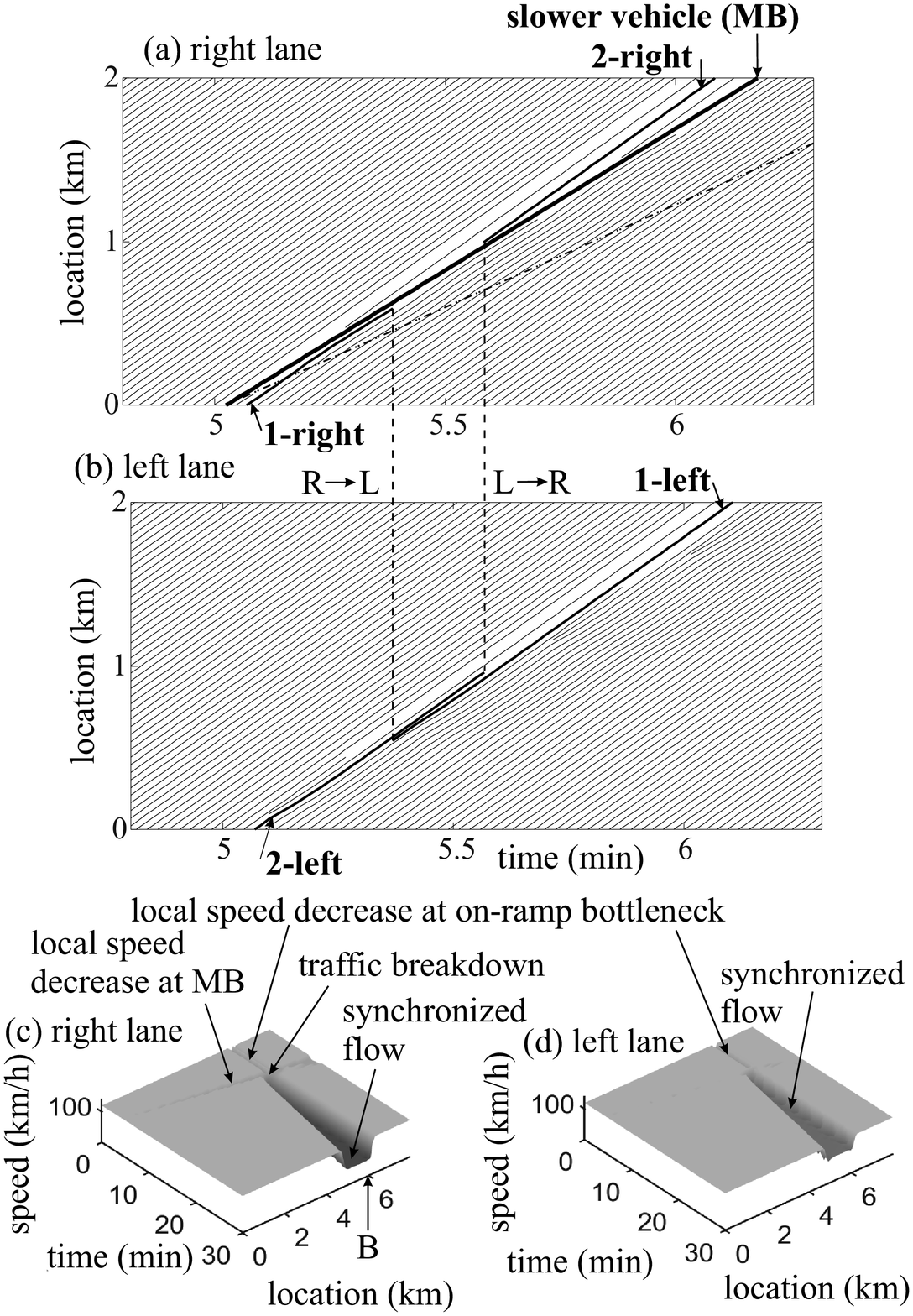}
\end{center}
\caption[]{Simulations of F$\rightarrow$S transition occurring due to downstream propagation of MB 
through the bottleneck   that   simplified version is shown in Fig.~\ref{MSP_MB_color} (b):
(a, b) Vehicle trajectories in the vicinity of MB in the right lane (a) and left lane (b). (c, d)
Speed in space and time in the right lane (c) and left lane
 (d). $q_{\rm in}=$ 2571
(vehicles/h)/lane, $q_{\rm on}=$ 720 vehicles/h, $L_{\rm M}=$ 0.3 km.  
Other model parameters are the same as those in Fig.~\ref{Free_Flow_Bottl}. 
}
\label{MB_Breakdown}
\end{figure}
 
To simulate a moving bottleneck (MB) in Fig.~\ref{MSP_MB_color} (b), we have assumed
that there is a single  automated vehicle moving in the right lane at a maximum free flow speed
$v_{\rm MB}$ that is  less than $v_{\rm free}$.
Already at $v_{\rm MB}=$ 110 km/h that is only 10 km/h less than
$v_{\rm free}$, the slower vehicle acts as the MB (Figs.~\ref{MSP_MB_color} (b) and~\ref{MB_Breakdown}).
We have also assumed that
through the use of cooperative-driving    automated vehicles   receive the
information about the location and  speed of the MB. Within a MB merging region   of length $L_{\rm M}$, each   vehicle moving in the right lane   changes to the left lane to pass the MB  if safety conditions (\ref{g_prec_ACC}) are satisfied
(e.g., see vehicle 1  
in Figs.~\ref{MB_Breakdown} (a, b));  
  lane-changing rules (\ref{RL}), (\ref{LR}) are not applied within the MB merging region~\footnote{This MB model is the same as that in~\cite{KKl2010MB}
used in a stochastic discrete microscopic model for human-driving traffic.}.
 Other vehicles for which   conditions (\ref{g_prec_ACC}) are not satisfied have to move at the velocity 
$v_{\rm MB}$ behind the MB (trajectories of these vehicles are within a region between the MB trajectory  
  and a dashed-dotted line in
 Fig.~\ref{MB_Breakdown} (a)).
Some  vehicles moving in the left lane, after they have passed the MB location, change back to 
the right lane where they can move at the   speed $v_{\rm free}$ (vehicle 2  
in Figs.~\ref{MB_Breakdown} (a, b)). 

The MB causes a  speed decrease localized at the MB
that moves at the speed 
$v_{\rm MB}$ (Fig.~\ref{MB_Breakdown} (c)).  As   in human-driving traffic (Fig.~\ref{Nucleation_Emp} (b)),
when the local speed decrease   at the MB reaches other local speed decrease  
at the on-ramp bottleneck, an additional short-time local speed decrease occurs at the bottleneck;
this  acts as a nucleus for
 traffic breakdown (F$\rightarrow$S transition) at the bottleneck 
(Figs.~\ref{MSP_MB_color} (b) and~\ref{MB_Breakdown}).

\section{Range of highway capacities at any time instant \label{Range_S}}

\subsection{Minimum and maximum highway capacities}

We have found that  at any time instant  the metastability of free flow in automated-driving 
vehicular traffic  on two-lane road with the bottleneck is realized within a flow rate range
\begin{equation}
C_{\rm min}\leq q_{\rm sum} < C_{\rm max},
\label{range_formula}
\end{equation}
where 
$q_{\rm sum}=2q_{\rm in} +   q_{\rm on}$ is the total flow rate across the road in free flow;
 $C_{\rm min}$ and $C_{\rm max}$ are, respectively, minimum and maximum highway capacities. The physics
of the capacity range (\ref{range_formula}) is that
within this capacity range   an F$\rightarrow$S transition can be induced at the bottleneck.
This result is in accordance with the three-phase traffic theory of human-driving traffic.

   \begin{figure}
\begin{center}
\includegraphics[width = 8 cm]{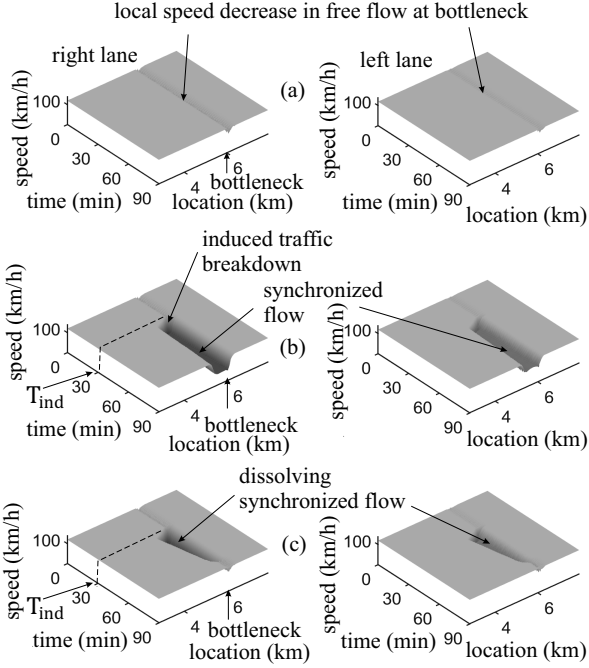}
\end{center}
\caption[]{Simulations of minimum capacity $C_{\rm min}=2q_{\rm in} +  q_{\rm on,   min}$ of free flow
  at bottleneck.
Speed in space and time in the right lane (left column) and in  left lane (right column) at different
 $q_{\rm on}$ at the same value $q_{\rm in}=$ 2571
(vehicles/h)/lane as that in Fig.~\ref{Free_Flow_Bottl}: (a) Free flow. (b) Induced traffic breakdown in (a).
(c) Induced dissolving synchronized flow. In
(a, b), $q_{\rm on}=q_{\rm on,   min}=$ 650 vehicles/h, i.e., $C_{\rm min}=$ 5792  vehicles/h; in
(c), $q_{\rm on}=$ 630 vehicles/h. 
In (b, c), as explained in Sec.~\ref{Free_S_Induced},   on-ramp inflow rate impulse has been applied; parameters of the impulse inducing either F$\rightarrow$S transition (induced traffic breakdown) (b) or
dissolving synchronized flow (c) at bottleneck are: $T_{\rm ind}=$ 30 min, $\Delta t=$ 5 min;
$\Delta q_{\rm on}=$ 250 vehicles/h in (b) and $\Delta q_{\rm on}=$ 270 vehicles/h in (c).
 Other model parameters are the same as those in Fig.~\ref{Free_Flow_Bottl}.  
}
\label{Min_Capacity} 
\end{figure}

  The   minimum   capacity $C_{\rm min}$ is explained in Fig.~\ref{Min_Capacity}:
At a given  $q_{\rm in}$,
there is a minimum on-ramp inflow rate denoted by $q_{\rm on}=q_{\rm on,   min}$
at which in an initial free flow   at the bottleneck (Fig.~\ref{Min_Capacity} (a)) an F$\rightarrow$S transition can still be induced (Fig.~\ref{Min_Capacity} (b)); the minimum capacity is equal to $C_{\rm min}=2q_{\rm in} +  q_{\rm on, min}$.
  At the model parameters, the  F$\rightarrow$S transition leads to the formation of a localized synchronized flow pattern (LSP)
at the bottleneck  (Fig.~\ref{Min_Capacity} (b)).
 Contrarily,  if  
\begin{equation}
q_{\rm sum} < C_{\rm min},
\label{min_less_formula}
\end{equation}
no F$\rightarrow$S transition  can be induced at the bottleneck:
synchronized flow induced at the bottleneck dissolves over time
(labeled by $\lq\lq$dissolving synchronized flow" in Fig.~\ref{Min_Capacity} (c)).

   \begin{figure}
\begin{center}
\includegraphics[width = 8 cm]{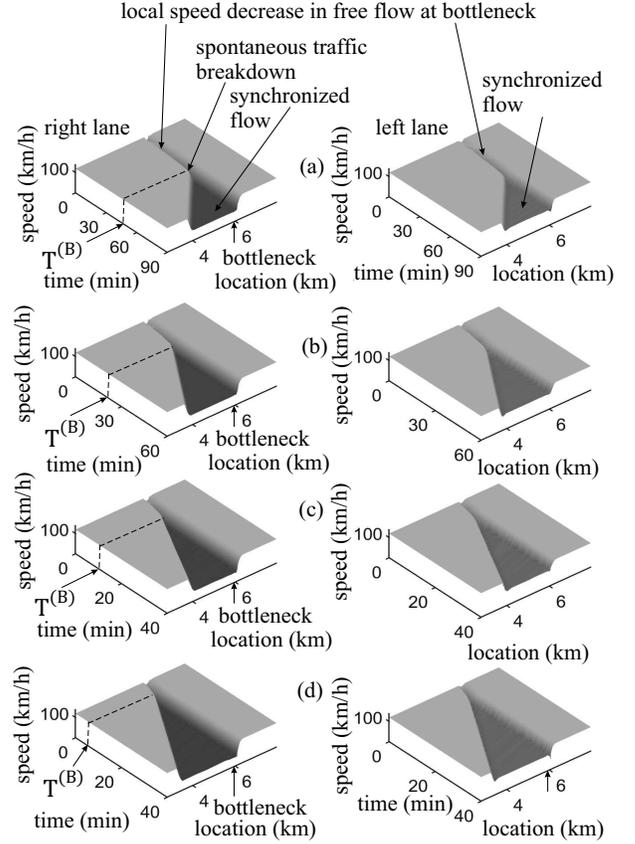}
\end{center}
\caption[]{Simulations of
spontaneous F$\rightarrow$S transition at bottleneck.
Speed in space and time in the right lane (left column) and in  left lane (right column) at different
 $q_{\rm on}$ at the same value $q_{\rm in}=$ 2571
(vehicles/h)/lane as that in Fig.~\ref{Free_Flow_Bottl}:
(a) $q_{\rm on}=$ 729 vehicles/h, 
$T^{\rm (B)}=$ 51 min.
(b) $q_{\rm on}=$ 740 vehicles/h, $T^{\rm (B)}=$ 19.8 min.
(c) $q_{\rm on}=$ 760 vehicles/h, $T^{\rm (B)}=$ 10 min.
(d) $q_{\rm on}=$ 780 vehicles/h, $T^{\rm (B)}=$ 5 min.  
Other model parameters are the same as those in Fig.~\ref{Free_Flow_Bottl}. 
}
\label{Spontaneous_F_S}
\end{figure}
 
 When the flow rate $q_{\rm sum}$ increases, a maximum highway capacity $C_{\rm max}$ can be reached.
The maximum   capacity $C_{\rm max}$ is a total flow rate $q_{\rm sum}$
that separates two qualitatively different phenomena: (i) When condition (\ref{range_formula}) is satisfied,
then free flow is in a metastable state with respect to the F$\rightarrow$S transition
at the bottleneck (Figs.~\ref{Induced_F_S_Bottl} and~\ref{Min_Capacity} (b)). (ii) When condition 
\begin{equation}
q_{\rm sum} > C_{\rm max} 
\label{max_larger_formula}
\end{equation}
is satisfied, then free flow is in an unstable state with respect to a {\it spontaneous} F$\rightarrow$S transition
at the bottleneck 
(Fig.~\ref{Spontaneous_F_S}).
At a given   flow rate $q_{\rm in}$,  
  the increase in $q_{\rm sum}$
is achieved through the increase in 
    $q_{\rm on}$. In this case,  the maximum capacity $C_{\rm max}$
	is reached, when the on-ramp inflow rate $q_{\rm on}$ is equal to some critical value denoted by
	$q_{\rm on}= q_{\rm on,   max}$, i.e.,   $C_{\rm max}=2 q_{\rm in}+ q_{\rm on,   max}$.

\subsection{Time delay of  spontaneous traffic breakdown (spontaneous F$\rightarrow$S transition) \label{T_B_Sec}}

 There is a time delay of the spontaneous F$\rightarrow$S transition at the bottleneck
denoted by $T^{\rm (B)}$
(Figs.~\ref{Spontaneous_F_S} and~\ref{TimeDelay_F_S}):    Under condition (\ref{max_larger_formula}), it has been found that the less the difference
$q_{\rm sum}-C_{\rm max}$, the longer the time delay $T^{\rm (B)}$ is (Fig.~\ref{TimeDelay_F_S}). 
In   the time-delay--flow-rate plane, condition $q_{\rm sum} = C_{\rm max}$
determines an asymptote (dashed vertical line in Fig.~\ref{TimeDelay_F_S})
that separates metastable free flow
(left of the asymptote) and unstable free flow   with respect to the F$\rightarrow$S transition
(right of the asymptote)~\footnote{It must be emphasized that we study a {\it deterministic} model of automated-driving vehicular traffic
of Sec.~\ref{Model_Sec},
 in which   {\it no random} local disturbances of speed, flow rate, or/and density
occur. Rather than through random effects, in such a $\lq\lq$deterministic  limit"  of the three-phase traffic theory,  local
 disturbances appear {\it only}
through   vehicle interactions with each other.
In particular, such vehicle interactions are caused by vehicle merging at the on-ramp bottleneck
as well as by lane-changing behavior. For this reason,  
if $q_{\rm sum} = C_{\rm max}$, then the traffic system is in a $\lq\lq$intermediate" free flow
state in which any (even very small) addition time-limited
 local speed decrease at the bottleneck causes
an F$\rightarrow$S transition at the bottleneck.
It has been proven that the smaller the time-limited additional local speed decrease at the bottleneck is,
the longer the time delay $T^{\rm (B)}$ of the F$\rightarrow$S transition at the bottleneck. 
}.

\begin{figure}
\begin{center}
\includegraphics[width = 8 cm]{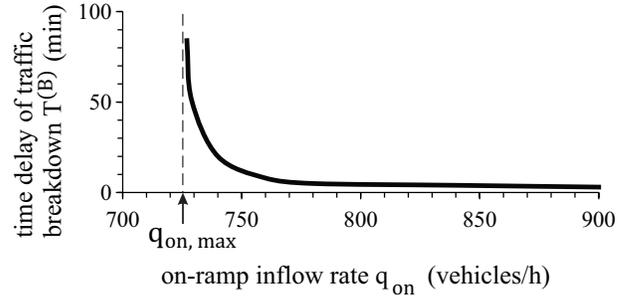}
\end{center}
\caption[]{Continuation of Fig.~\ref{Spontaneous_F_S}.
Dependence of the time delay $T^{\rm (B)}$ of
spontaneous F$\rightarrow$S transition at bottleneck on on-ramp inflow rate $q_{\rm on}$
at the given flow rate $q_{\rm in}=$ 2571
(vehicles/h)/lane. Calculated values $q_{\rm on,   max}=$ 726
vehicles/h, $C_{\rm max}=2 q_{\rm in}+ q_{\rm on,   max}=$ 5868 vehicles/h.
}
\label{TimeDelay_F_S}
\end{figure}

\subsection{Range of discontinuity of over-acceleration rate}

Within the flow-rate range (\ref{range_formula}) there can be either a free flow state or a synchronized flow state
at the bottleneck. Under condition $q_{\rm in}=$const, the range (\ref{range_formula}) is equivalent to the on-ramp inflow-rate range 
 (Fig.~\ref{Over-Accl_Rate-flow-rate})
\begin{equation}
q_{\rm on,   min} \leq q_{\rm on}<q_{\rm on,   max}.
\label{q_on_min_q_on_max}
\end{equation}
When the initial state is   free flow and  $q_{\rm on}$ increases, then
at $q_{\rm on} >q_{\rm on,   max}$ a spontaneous F$\rightarrow$S transition occurs  with a delay time
$T^{\rm (B)}$
(Figs.~\ref{Spontaneous_F_S} and~\ref{TimeDelay_F_S}).
The emergent synchronized flow persists due to
the discontinuity in the over-acceleration rate (Sec.~\ref{Adaptation_Over_Sec}):
The over-acceleration rate decreases sharply  (down-arrow in Fig.~\ref{Over-Accl_Rate-flow-rate} (a)), respectively,
the mean time delay in over-acceleration increases sharply
(up-arrow in Fig.~\ref{Over-Accl_Rate-flow-rate} (b)).

\begin{figure}
\begin{center}
\includegraphics[width = 8 cm]{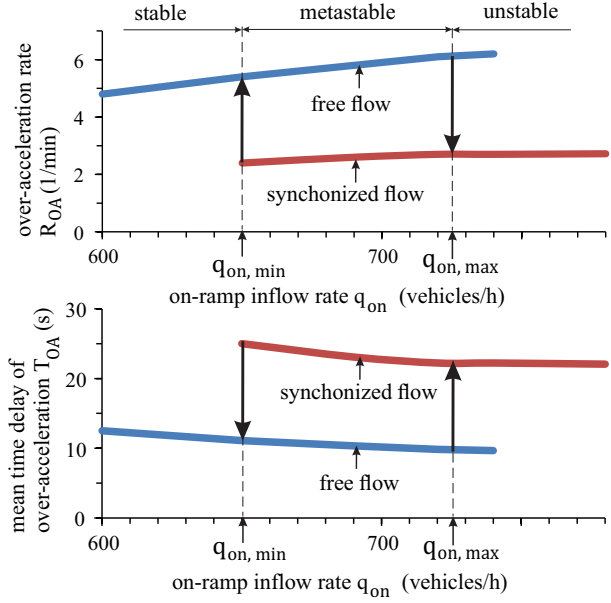}
\end{center}
\caption[]{Simulated range of the discontinuity in over-acceleration rate (a) and in   mean time-delay in over-acceleration (b) as functions of the on-ramp inflow rate $q_{\rm on}$ at   given flow rate $q_{\rm in}=$ 2571
(vehicles/h)/lane: Z-characteristics of the F$\rightarrow$S transition in automated-driving vehicular traffic on two-lane road with bottleneck.
 Other model parameters are the same as those in Fig.~\ref{Free_Flow_Bottl}.
}
\label{Over-Accl_Rate-flow-rate}
\end{figure}

When  $q_{\rm on}$ decreases,
 synchronized flow exists in the  range (\ref{q_on_min_q_on_max}).
Only when $q_{\rm on}$ becomes less than $q_{\rm on,  min}$,
a return spontaneous S$\rightarrow$F transition occurs at the bottleneck; respectively,
free flow recovers at the bottleneck.
Thus, there is a Z-characteristic for traffic breakdown at the bottleneck
that shows stable, metastable, and unstable states of free flow  with respect to
the  F$\rightarrow$S transition at the bottleneck
(Fig.~\ref{Over-Accl_Rate-flow-rate}).

\subsection{Physics of   spontaneous traffic breakdown \label{Physics_Spon_F_S_Sec}}

The spontaneous F$\rightarrow$S transition occurs at    $t=T^{\rm (B)}$ (Sec.~\ref{T_B_Sec}) when 
a {\it sequence of two   R$\rightarrow$L lane-changing} occurs: One of them is realized at the upstream front
 of the local speed decrease (vehicle 1 in Figs.~\ref{Spontaneous_F_S_Bottl_tr2} (a, b)) and another 
 occurs at its downstream front (vehicle 2 in Figs.~\ref{Spontaneous_F_S_Bottl_tr2} (a, b)).
Simultaneously, 
a  drop in the over-acceleration rate $R_{\rm OA}$ (Fig.~\ref{Spontaneous_F_S_Bottl_tr2} (c))
and, respectively, a jump in the mean time delay in  over-acceleration   $T_{\rm OA}$ are realized
 (Fig.~\ref{Spontaneous_F_S_Bottl_tr2} (d)). As explained in Sec.~\ref{Adaptation_Over_Sec}, this discontinuous 
behavior of over-acceleration causes the abrupt transformation of 
  the local speed decrease   in free flow
 at the bottleneck  into synchronized flow. The boundaries of synchronized flow are
given by the upstream synchronized flow front propagating upstream (dashed curve $\lq\lq$S-up" in 
Fig.~\ref{Spontaneous_F_S_Bottl_tr2} (a, b))
and  the downstream synchronized flow front   (dashed-dotted curve $\lq\lq$S-down") fixed at the bottleneck.

 \begin{figure}
\begin{center}
\includegraphics[width = 8 cm]{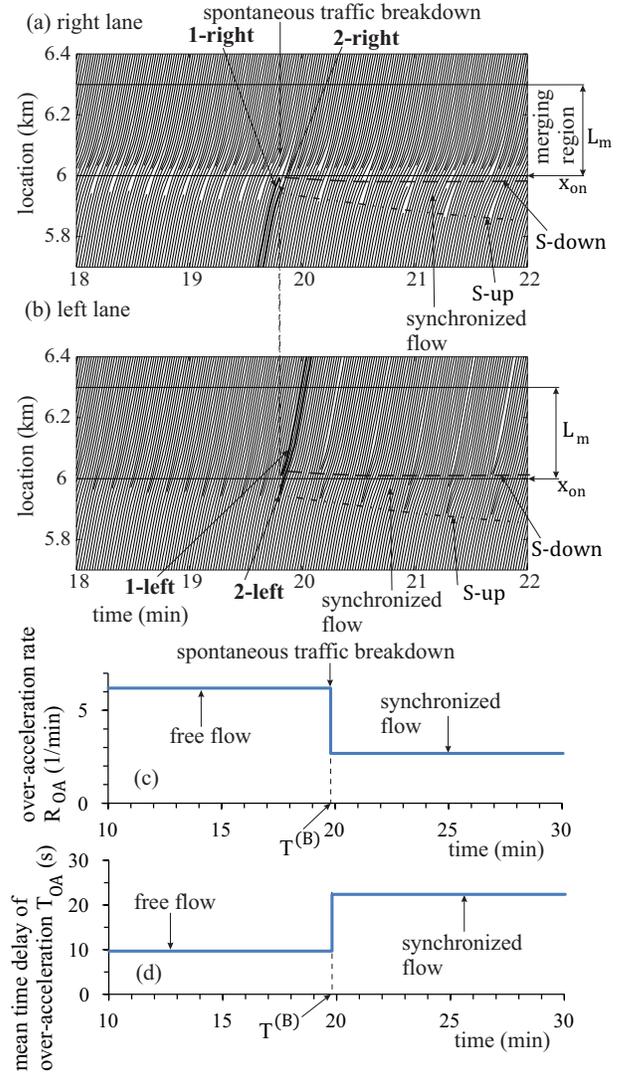}
\end{center}
\caption[]{Continuation of Fig.~\ref{Spontaneous_F_S} (b).
Features of spontaneous traffic breakdown:
(a, b) Vehicle trajectories in the right lane (a) and left lane (b).
(c, d) Time-dependencies of the averaged over-acceleration rate $R_{\rm OA}$ (c)  
and  the mean time delay in  over-acceleration   $T_{\rm OA}$ (d).    
}
\label{Spontaneous_F_S_Bottl_tr2}
\end{figure}

\begin{figure}
\begin{center}
\includegraphics[width = 8 cm]{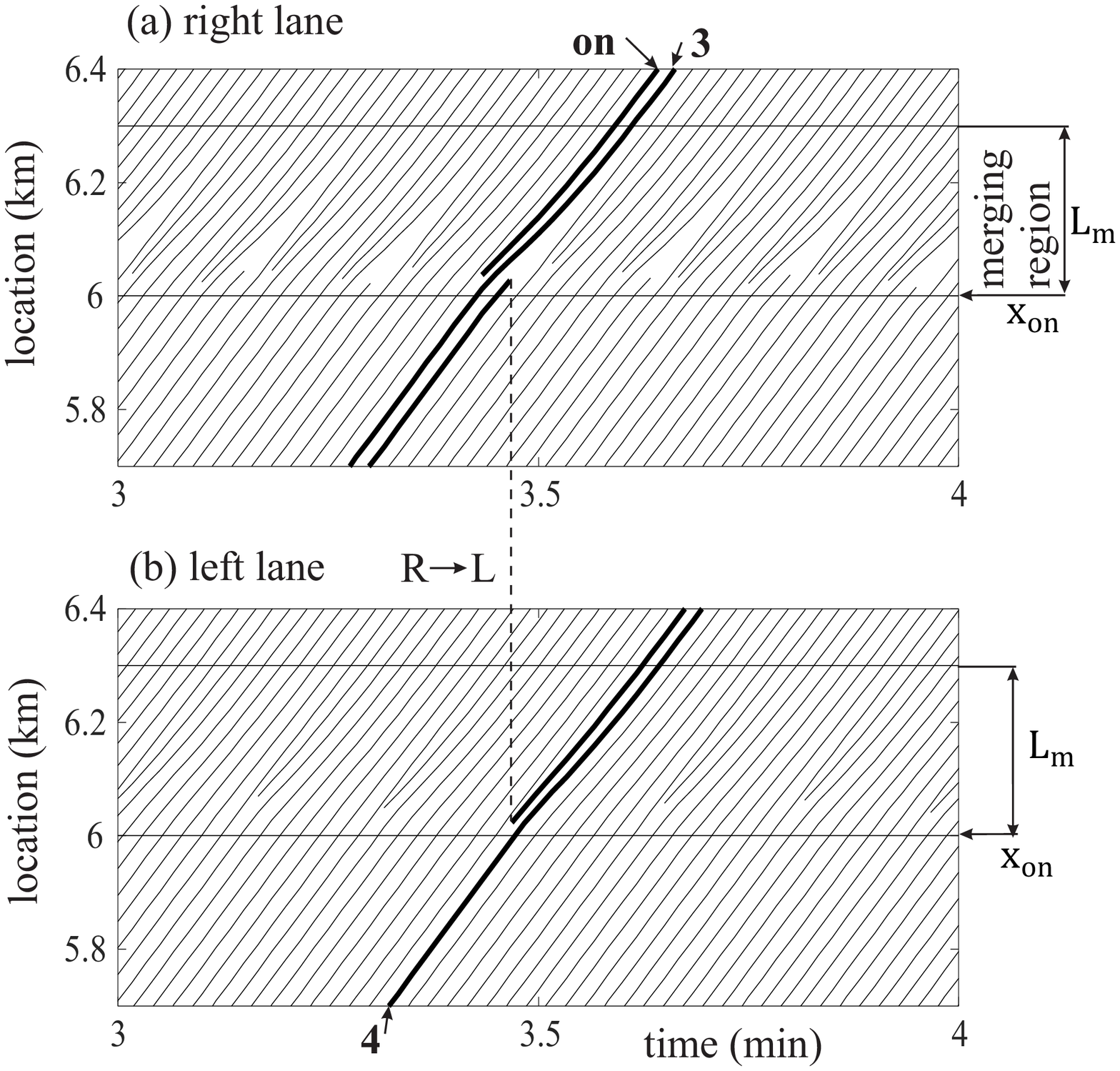}
\end{center}
\caption[]{Continuation of Fig.~\ref{Spontaneous_F_S} (b):
Vehicle trajectories in free flow at the bottleneck
at   $t\ll T^{\rm (B)}$. 
}
\label{Spontaneous_F_S_Bottl_tr_free1}
\end{figure}

 \begin{figure}
\begin{center}
\includegraphics[width = 8 cm]{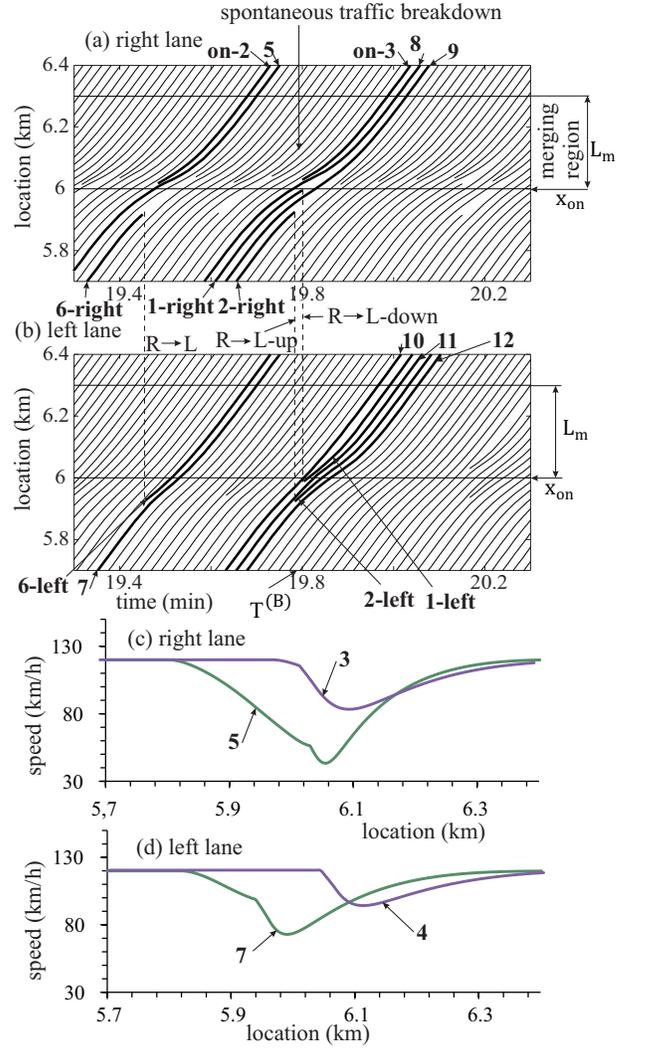}
\end{center}
\caption[]{Continuation of Fig.~\ref{Spontaneous_F_S} (b).
(a, b)   Vehicle trajectories   $\lq\lq$on-2", 5--7 are in free flow  
  at time that is   about 30 s less than $t=T^{\rm (B)}$;  trajectories $\lq\lq$on-3", 8--12
are related to the   time of traffic breakdown (F$\rightarrow$S transition)
        $t= T^{\rm (B)}$. 
(c, d) Comparison of location-functions of  speeds  
 for vehicles 3 and 4 taken from Fig.~\ref{Spontaneous_F_S_Bottl_tr_free1} with
speeds on trajectories 5 and 7 from (a, b). 
Vehicles 1 and 2 are, respectively, the same as that in Fig.~\ref{Spontaneous_F_S_Bottl_tr2} (a, b).
Sequence of two R$\rightarrow$L   
lane-changing effects of vehicles 1 and 2 that causes spontaneous traffic breakdown
are labeled by R$\rightarrow$L-down and R$\rightarrow$L-up, respectively.
}
\label{Spontaneous_F_S_Bottl_tr_fine}
\end{figure}

 \begin{figure}
\begin{center}
\includegraphics[width = 8 cm]{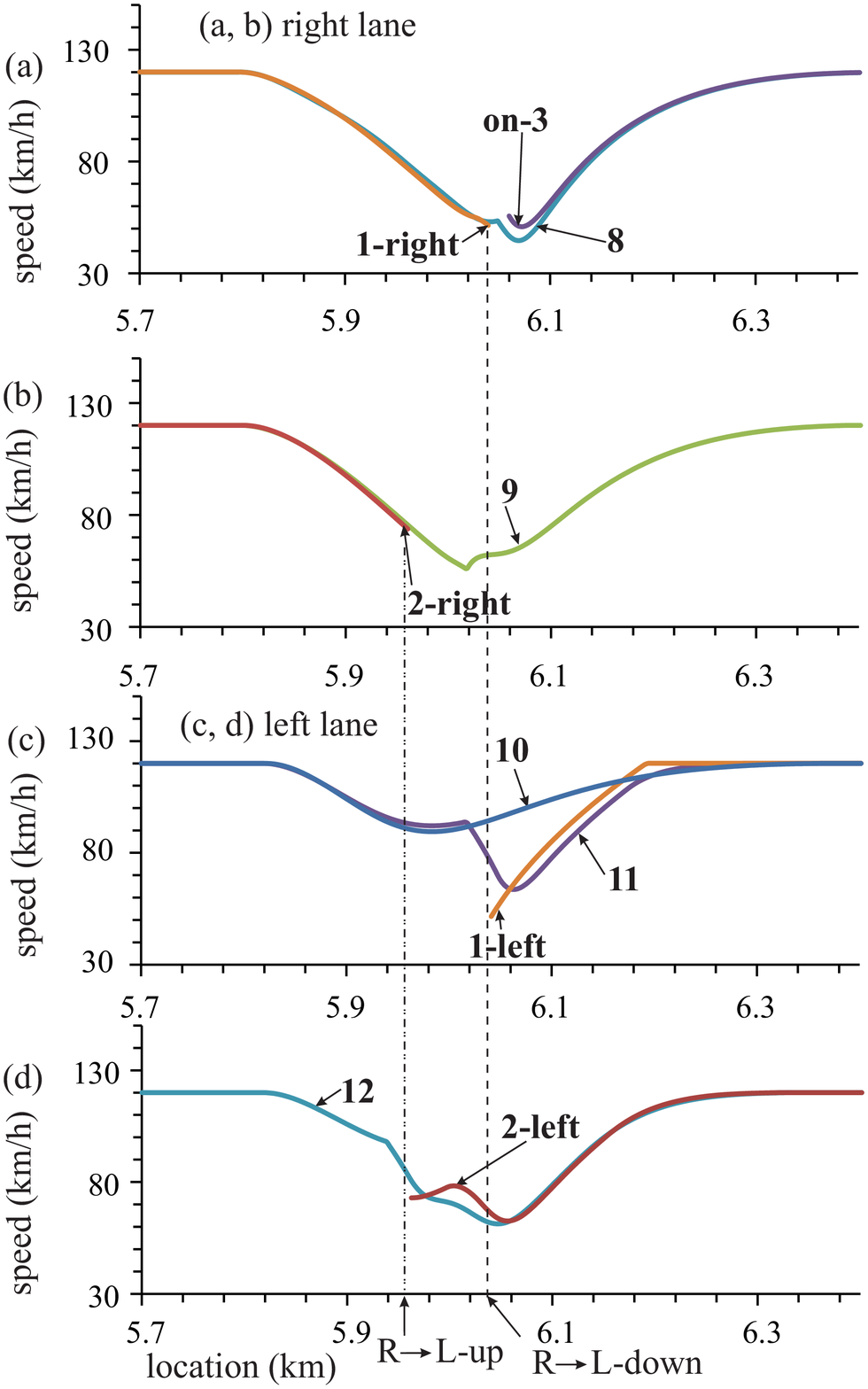}
\end{center}
\caption[]{Continuation of Fig.~\ref{Spontaneous_F_S_Bottl_tr_fine}:
Location-functions of  speed for some vehicles whose numbers are the same as that in Fig.~\ref{Spontaneous_F_S_Bottl_tr_fine},
respectively; vehicles 1 and 2 are, respectively, the same as that in Figs.~\ref{Spontaneous_F_S_Bottl_tr2} (a, b).
}
\label{Spontaneous_F_S_Bottl_tr_fine2}
\end{figure}

The physics of the maximum capacity $C_{\rm max}$ and time delay $T^{\rm (B)}$  of spontaneous
traffic breakdown  is as follows. As found,
at $q_{\rm on}< q_{\rm on, max}$ the minimum speed within the local speed decrease in free flow 
at the bottleneck does not almost depend on time (Fig.~\ref{Free_Flow_Bottl}).
Contrarily, at $q_{\rm on}> q_{\rm on, max}$ (Fig.~\ref{Spontaneous_F_S}), the 
 minimum speed within the local speed decrease in free flow grows continuously over time. Indeed,
at $t\ll T^{\rm (B)}$ (Fig.~\ref{Spontaneous_F_S_Bottl_tr_free1})
  minimum speeds of vehicles 3 and 4 
are considerably larger than  minimum speeds, respectively,  of vehicles 5 and 7  moving
in free flow at time that is only about 30 s less than $t=T^{\rm (B)}$ 
  (Figs.~\ref{Spontaneous_F_S_Bottl_tr_fine} (c, d)). Thus, the maximum capacity $C_{\rm  max}$ separates free flow states  
at $q_{\rm on}<q_{\rm on, max}$,
in which the local speed decrease at the bottleneck does not growth over time,  from
free flow states at $q_{\rm on}>q_{\rm on, max}$, in which the local speed decrease does continuously grow  over time.

 The continuous reduction of the minimum speed within
 the local speed decrease in free flow at the bottleneck over time
 has to have a limit  that can be considered a critical minimum speed:
After   vehicle $\lq\lq$on-3" has merged from the on-ramp,   minimum speeds 
of vehicles 8 and 9 moving in the right lane become   low enough; this causes    
  the sequence of two R$\rightarrow$L   
lane-changing of vehicles 1 and 2  (Figs.~\ref{Spontaneous_F_S_Bottl_tr_fine2} (a, b)).
Slow vehicles 1-left and 2-left (Figs.~\ref{Spontaneous_F_S_Bottl_tr_fine2} (c, d)) force the following
vehicles 11 and   12 moving in the left lane to decelerate strongly.
At so low speed in the left lane  
the over-acceleration rate $R_{\rm OA}$ drops and, respectively,
the mean time delay in over-acceleration increases sharply (Figs.~\ref{Spontaneous_F_S_Bottl_tr2} (c, d)); as a result,
the speed adaptation overcomes the over-acceleration.  

It takes some time   for the continuous reduction  of the minimum speed within the   local speed decrease  to 
the critical speed  in free flow  at which traffic breakdown occurs at the bottleneck. This time interval determines
  time delay $T^{\rm (B)}$ of traffic breakdown (F$\rightarrow$S transition).
We have found  that the more the on-ramp inflow rate  $q_{\rm on}$ exceeds
the critical value $q_{\rm on, max}$, the quicker the critical minimum speed  in free flow at the bottleneck
is reached. This explains   the decreasing character of function $T^{\rm (B)}(q_{\rm on})$ 
(Fig.~\ref{TimeDelay_F_S}).

	\section{Generalization of   nucleation features     of   F$\rightarrow$S transition in 
	automated-driving traffic \label{Generalization_Sec}}
	
	Up to now we have used only one chosen set of model parameters,
	  to demonstrate that automated-driving traffic does exhibit the basic feature of the three-phase traffic theory --
	  the  nucleation character of  an F$\rightarrow$S transition at the  bottleneck. To disclose the physics of
		this F$\rightarrow$S transition, we have studied its features
	under a change in the on-ramp inflow rate  $q_{\rm on}$ at the bottleneck   
	(Secs.~\ref{Meta_S} and~\ref{Range_S}). However, do   basic results of this paper
	about the  nucleation character of  the F$\rightarrow$S transition at the  bottleneck and the existence of a range of highway capacities
 	 remain in automated-driving vehicular traffic, when   model parameters are changed?
	
	\subsection{Effect of lane-changing model parameters on   F$\rightarrow$S transition \label{Param_lane-changing_Sec}}

\begin{figure}
\begin{center}
\includegraphics[width = 8 cm]{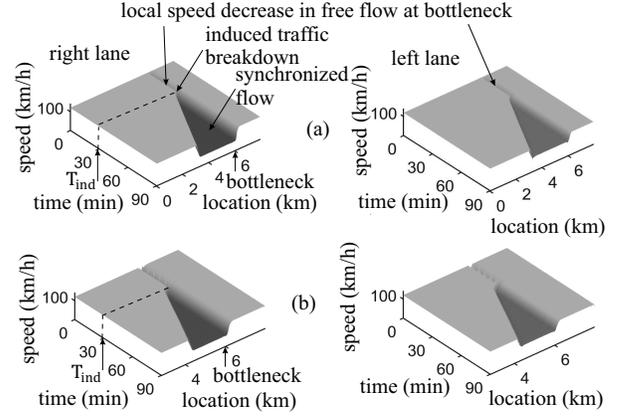}
\end{center}
\caption[]{ Speed in space and time in the right lane (left column) and in left lane (right column)
at the same flow rate $q_{\rm in}=$ 2571
(vehicles/h)/lane as that in Fig.~\ref{Free_Flow_Bottl}.  (a, b) Induced S$\rightarrow$J transition
that has been simulated as that   in Fig.~\ref{Induced_F_S_Bottl}.
(a) Symmetric lane-changing parameters $\delta_{1}=\delta_{2}=$ 1 m/s in (\ref{RL}), (\ref{LR}),
  $q_{\rm on}=$ 720 vehicles/h, $\Delta q_{\rm on}=$ 180 vehicles/h.
(b) Symmetric safety parameters $\tau_{1}=\tau_{2}=$ 0.4 s 
in (\ref{g_prec_ACC}),
  $q_{\rm on}=$ 700 vehicles/h,
	$\Delta q_{\rm on}=$ 200 vehicles/h.     $T_{\rm ind}=$ 30 min,
  $\Delta t=$ 2 min.
Other model parameters are the same as those in Fig.~\ref{Free_Flow_Bottl}. 
}
\label{Generalization_F_S}
\end{figure}

	We have found that as long as
  new model parameters  in lane-changing rules (\ref{RL})--(\ref{g_prec_ACC})
   enable a   distribution  of on-ramp inflow between road lanes in free flow,    all qualitative results
presented above	remain the same ones. Examples are shown in Fig.~\ref{Generalization_F_S}
for symmetric lane-changing parameters $\delta_{1}=\delta_{2}$ in (\ref{RL}), (\ref{LR})
(Fig.~\ref{Generalization_F_S} (a)) and for symmetric safety parameters $\tau_{1}=\tau_{2}$  
in (\ref{g_prec_ACC}).

\subsection{Diagrams of F$\rightarrow$S transition at bottleneck \label{Diagram_Sec}}

\begin{figure}
\begin{center}
\includegraphics[width = 8 cm]{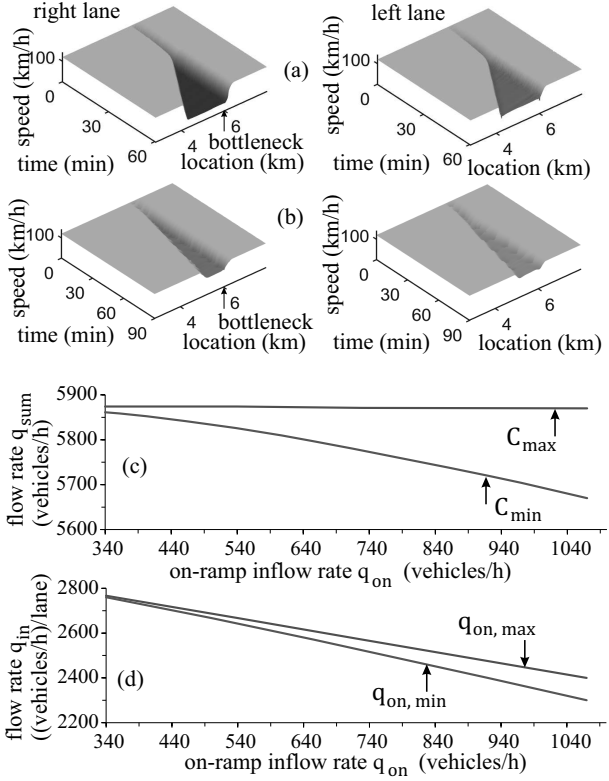}
\end{center}
\caption[]{Simulations
 of the nucleation nature of the F$\rightarrow$S transition at the bottleneck  
 for different  values $q_{\rm in}$:
(a, b) 
Speed in space and time in the right lane (left column) and in   left lane (right column) for spontaneous  F$\rightarrow$S transition:
(a) $q_{\rm in}=$ 2449 (vehicles/h)/lane, $q_{\rm on}=$ 980 vehicles/h, $T^{\rm (B)}=$ 26 min.
(b) $q_{\rm in}=$ 2769 (vehicles/h)/lane, $q_{\rm on}=$ 340 vehicles/h, $T^{\rm (B)}=$ 24 min.
(c) Dependencies of minimum highway capacity
$C_{\rm min}$ and maximum highway capacity $C_{\rm max}$
  on $q_{\rm on}$.
(d) Dependencies    $q_{\rm in}(q_{\rm on})$ related to $C_{\rm min}(q_{\rm on})$
(curve denoted by  $q_{\rm on, min}$) and $C_{\rm max}(q_{\rm on})$ (curve denoted by $q_{\rm on, max}$), respectively.
Other model parameters are the same as those in Fig.~\ref{Free_Flow_Bottl}. 
}
\label{Different_q_in_Fig}
\end{figure}

To understand  the nucleation nature of the F$\rightarrow$S transition 
in automated-driving traffic, up to now we have used only one   given flow rate in free flow upstream of the bottleneck
$q_{\rm in}=$ 2571
(vehicles/h)/lane. We have found that 
  the nucleation nature of the F$\rightarrow$S transition at the bottleneck  remains   
when $q_{\rm in}$ changes 
(Fig.~\ref{Different_q_in_Fig}). In particular,
  maximum capacity $C_{\rm max}$
does not almost depend on   $q_{\rm on}$,
 whereas  
 minimum capacity $C_{\rm min}$ is a decreasing   function of $q_{\rm on}$:
   the larger the on-ramp inflow rate $q_{\rm on}$, the larger the capacity range
	$C_{\rm max}-C_{\rm min}$
	(Fig.~\ref{Different_q_in_Fig} (c)). 
 When the flow rate $q_{\rm on}$ increases, the  flow-rate range   $q_{\rm on, max}-q_{\rm on, min}$, within which
free flow is metastable with respect to the F$\rightarrow$S transition at the bottleneck, increases
 (Fig.~\ref{Different_q_in_Fig} (d)).

 \begin{figure}
\begin{center}
\includegraphics[width = 8 cm]{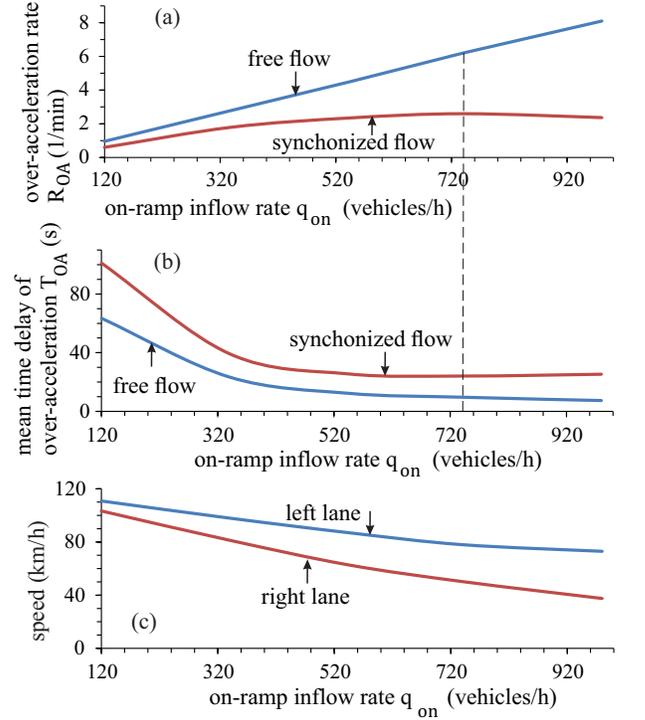}
\end{center}
\caption[]{Characteristics of spontaneous F$\rightarrow$S transition  at different flow rates $q_{\rm in}$
in free flow upstream of bottleneck.
(a, b) The discontinuity of the over-acceleration rate:
On-ramp inflow-rate dependence of the over-acceleration rate $R_{\rm OA}$ (a) and the mean time delay in over-acceleration
$T_{\rm OA}$ (b) in initial free flow (curves $\lq\lq$free flow") and in synchronized flow (curves $\lq\lq$synchronized
 flow") that has occurred due to 
F$\rightarrow$S transition. (c) Synchronized flow speeds occurring at the bottleneck after the F$\rightarrow$S transition in the right and left lanes.  In (a, b),  on-ramp inflow-rates
$q_{\rm on}$ on the x-axis  are   slightly exceed corresponding values $q_{\rm on, max}$
(we have used $q_{\rm on}=q_{\rm on, max}+ \delta q$, where parameter $\delta q=$ 14 vehicles/h)
 calculated by different values $q_{\rm in}$;
for explanations of (a, b),   a dashed vertical line related to $q_{\rm on}=$ 740 vehicles/h    
 has been drawn to show, respectively, the same values $R_{\rm OA}$ and $T_{\rm OA}$ at curves
$R_{\rm OA}(q_{\rm on})$ and $T_{\rm OA}(q_{\rm on})$ as those
in Figs.~\ref{Spontaneous_F_S_Bottl_tr2} (c, d) for $q_{\rm in}=$ 2571
(vehicles/h)/lane.
Other model parameters are the same as those in Fig.~\ref{Free_Flow_Bottl}. 
}
\label{Diagram_Fig}
\end{figure}	 

 At any value $q_{\rm in}$, at which the F$\rightarrow$S transition can occur,
the physics of the F$\rightarrow$S transition is qualitatively the same as that 
disclosed in Secs.~\ref{Meta_S}
and~\ref{Range_S}. In particular,
the nature of the F$\rightarrow$S transition is caused by the discontinuity of the over-acceleration rate (Fig.~\ref{Diagram_Fig} (a, b))
as well as its competition with speed adaptation.
Features of synchronized flow occurring due to the F$\rightarrow$S transition
(Sec.~\ref{Effect_Dis_Sec}) remain also the same when   $q_{\rm in}$  changes.
The  speeds in synchronized flow      in the right and left lanes
at the bottleneck are decreasing
functions of the on-ramp inflow rate 
(Fig.~\ref{Diagram_Fig} (c)).

  \subsection{Lane-asymmetric nucleation of F$\rightarrow$S transition  \label{Lane-asym_S}}

	Because the nucleation nature of F$\rightarrow$S transition
	in automated-driving traffic at the bottleneck is determined by the existence of the discontinuity in
	R$\rightarrow$L  lane-changing rate
	(Sec.~\ref{Meta_S}), a question can arise:
	Does the nucleation nature of F$\rightarrow$S transition remain if the lane-changing rules
are	changed  qualitatively? Indeed, as known, cooperative driving in automated-driving traffic
 could permit the realization of  different
	lane-changing rules that
 enable a   distribution  of on-ramp inflow between road lanes in free flow
as done through   lane-changing rules (\ref{RL})--(\ref{g_prec_ACC}).   
 In  (\ref{RL})--(\ref{g_prec_ACC}),
	at a large speed difference between lanes no speed limitation for  lane-changing  has been assumed. When a 
	vehicle moving at a slow speed $v(t)$ changes from the right lane to the left lane, the vehicle can force the following vehicle
	moving at a larger speed $v^{-}(t)$ to decelerate
	strongly. This can considerably decrease comfortable driving and sometimes traffic safety. 

 \begin{figure}
\begin{center}
\includegraphics[width = 8 cm]{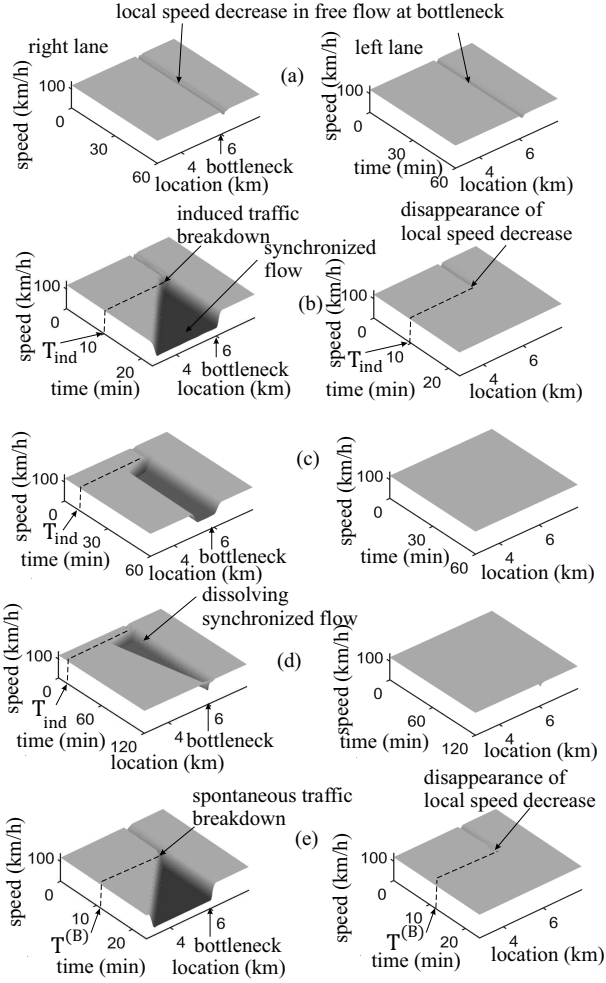}
\end{center}
\caption[]{Simulations of asymmetric
  F$\rightarrow$S transition at bottleneck that occurs in model of Sec.~\ref{Model_Sec}, when, in addition to safety conditions
(\ref{g_prec_ACC}),	 condition (\ref{speed_diff}) is used.
Speed in space and time in the right lane (left column) and in  left lane (right column) at different
 $q_{\rm on}$ at the same flow rate $q_{\rm in}=$ 2571
(vehicles/h)/lane as that in Fig.~\ref{Free_Flow_Bottl}:
(a) Local speed decrease at bottleneck in free flow, $q_{\rm on}=$ 720 vehicles/h. 
(b) Induced F$\rightarrow$S transition in free flow of (a); parameters of on-ramp inflow impulse: 
$T_{\rm ind}=$ 10 min, $\Delta q_{\rm on}=$ 180 vehicles/h, $\Delta t=$ 1 min.
(c) Induced F$\rightarrow$S transition at $q_{\rm on}=q_{\rm on, min}=$ 360 vehicles/h; 
$T_{\rm ind}=$ 10 min, $\Delta q_{\rm on}=$ 540 vehicles/h, $\Delta t=$ 2 min.
(d) Dissolving synchronized flow at $q_{\rm on}=$ 350 vehicles/h that is less
than $q_{\rm on, min}$; 
$T_{\rm ind}=$ 10 min, $\Delta q_{\rm on}=$ 550 vehicles/h, $\Delta t=$ 2 min.
  (e) Spontaneous F$\rightarrow$S transition at $q_{\rm on}=$ 727 vehicles/h that is larger
than $q_{\rm on, max}=$ 724 vehicles/h; 
$T^{\rm (B)}=$ 11.5 min. In (\ref{speed_diff}), $T_{\rm p}=$ 3.3 s, $g_{\rm p}=$ 2 m.
Other model parameters are the same as those in Fig.~\ref{Free_Flow_Bottl}. 
}
\label{F_S_asym}
\end{figure}

Cooperative driving
	  can solve this   problem through some  safety condition 
\begin{equation}
g^{-}(t) + (v(t)-v^{-}(t)) T_{\rm p} > g_{\rm p}
\label{speed_diff}
\end{equation}  
 used in addition to (\ref{g_prec_ACC}). Safety condition  (\ref{speed_diff}), in which
 $T_{\rm p}$ and $g_{\rm p}$  are constant parameters,
  limits R$\rightarrow$L lane-changing, when 
speed difference $v(t)-v^{-}(t)$ is large enough, whereas the space gap between these vehicles is not large enough for comfortable driving.

\subsubsection{Characteristics of lane-asymmetric nucleation of F$\rightarrow$S transition  }

Condition (\ref{speed_diff}) does not affect on R$\rightarrow$L lane-changing
in {\it free} flow (Fig.~\ref{F_S_asym} (a)): The same lane-changing rate is realized and the same local speed decrease
appears at the bottleneck as that  in Fig.~\ref{Free_Flow_Bottl}. There is  free flow metastability
with respect to the F$\rightarrow$S transition at the bottleneck as found in Secs.~\ref{Meta_S}
and~\ref{Range_S}; 
  condition (\ref{range_formula}) is also valid.
Moreover, values  $q_{\rm on, max}$ and, respectively,    $C_{\rm max}=2q_{\rm in}+q_{\rm on, max}$, which separate
metastable free flow from unstable free flow with respect to the F$\rightarrow$S transition at the bottleneck remains almost the same
  (Figs.~\ref{F_S_asym}--\ref{Charac_F_S_asym}).

However, the use of condition (\ref{speed_diff}) changes basically
the result of the F$\rightarrow$S transition  at the bottleneck:
In Sec.~\ref{Meta_S}, after the F$\rightarrow$S transition has occurred, 
synchronized flow emerges both in the right and in  left lanes
(Figs.~\ref{Induced_F_S_Bottl} and~\ref{Spontaneous_F_S}).
Contrarily, under condition (\ref{speed_diff}) 
the F$\rightarrow$S transition causes synchronized flow emergence in the right lane {\it only}
(Figs.~\ref{F_S_asym} (b--e)). For this reason, we can call the F$\rightarrow$S transition as
an {\it asymmetric}   F$\rightarrow$S transition at the bottleneck.

 \begin{figure}
\begin{center}
\includegraphics[width = 8 cm]{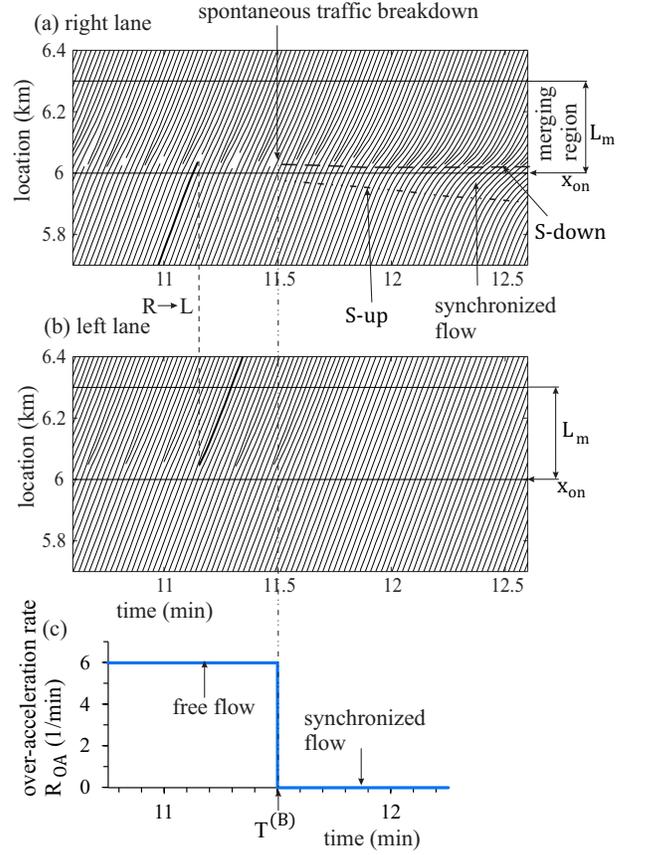}
\end{center}
\caption[]{Continuation of Fig.~\ref{F_S_asym} (e):
Features of asymmetric spontaneous traffic breakdown:
(a, b) Vehicle trajectories in the right lane (a) and left lane (b).
(c) Time-dependencies of the averaged over-acceleration rate $R_{\rm OA}$.  
}
\label{Spontaneous_F_S_Bottl_tr_asym}
\end{figure} 

Moreover, after 
 the asymmetric   F$\rightarrow$S transition has occurred
{\it no} local speed decrease remains in free flow in the left lane at the bottleneck
(right column in Figs.~\ref{F_S_asym} (b--e)).
 The disappearance of the local speed decrease  in free flow in the left lane at the bottleneck
is explained by the drop in the R$\rightarrow$L lane-changing rate to zero 
 during the asymmetric   F$\rightarrow$S transition
(Fig.~\ref{Spontaneous_F_S_Bottl_tr_asym}): No R$\rightarrow$L lane-changing
is realized at $t>T^{\rm (B)}$, i.e., after the asymmetric   F$\rightarrow$S transition 
 has occurred  at $t=T^{\rm (B)}$
(Fig.~\ref{Spontaneous_F_S_Bottl_tr_asym} (a, b)). Respectively, there is a drop
in the over-acceleration rate $R_{\rm OA}$ from the rate $R_{\rm OA}$
 in free flow  to $R_{\rm OA}=$ 0 in synchronized flow
(Fig.~\ref{Spontaneous_F_S_Bottl_tr_asym} (c); one of these R$\rightarrow$L lane-changing at $t<T^{\rm (B)}$
is marked by dashed vertical line R$\rightarrow$L in Figs.~\ref{Spontaneous_F_S_Bottl_tr_asym} (a, b))~\footnote{The rate $R_{\rm OA}$
 in free flow is nearly the same   as that  when   condition (\ref{speed_diff})
is not used (Secs.~\ref{Meta_S}--\ref{Diagram_Sec}).}. 
The physics of   this effect is as follows.
When synchronized flow begins to emerge in the right lane, the speed difference $v(t)-v^{-}(t)$ in (\ref{speed_diff})
becomes large enough. This prevents the R$\rightarrow$L lane-changing.

\begin{figure}
\begin{center}
\includegraphics[width = 8 cm]{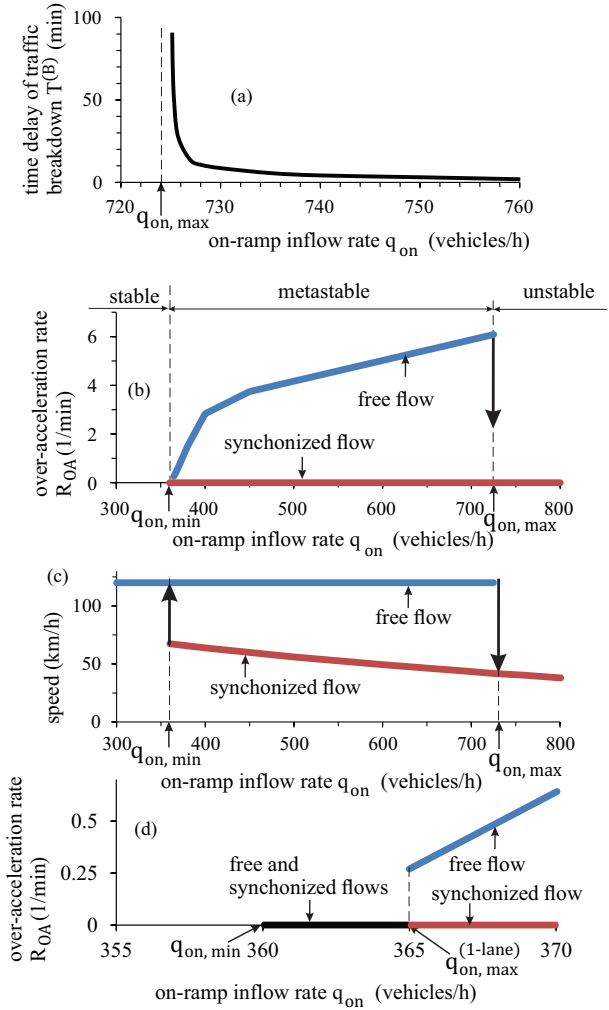}
\end{center}
\caption[]{Simulated characteristics of asymmetric F$\rightarrow$S transition
on two-lane road with bottleneck  
at the same   value $q_{\rm in}=$ 2571
(vehicles/h)/lane as that in Figs.~\ref{Free_Flow_Bottl}--\ref{Speed-Time-lane-symmetric}.
(a) Dependence of   time delay $T^{\rm (B)}$ of spontaneous traffic breakdown
on   $q_{\rm on}$;   $q_{\rm on,   max}=$ 724
vehicles/h, $C_{\rm max}=2 q_{\rm in}+ q_{\rm on,   max}=$ 5868 vehicles/h. (b, c) Simulated Z-characteristics of the
asymmetric F$\rightarrow$S transition:
 The discontinuity in over-acceleration rate (b) and speed (c) as functions of  $q_{\rm on}$.
(d) A small part of (b) in a large scale in vicinity of $q_{\rm on}=q_{\rm on,   min}$. Other model parameters are the same as those in Fig.~\ref{F_S_asym}.
}
\label{Charac_F_S_asym}
\end{figure}

We have found that as in  Fig.~\ref{TimeDelay_F_S},  time delay $T^{\rm (B)}$ of spontaneous asymmetric F$\rightarrow$S transition that occurs at $q_{\rm on}>q_{\rm on,   max}$ is also a strongly falling on-ramp inflow-rate function (Fig.~\ref{Charac_F_S_asym} (a)).
However, under the asymmetric F$\rightarrow$S transition there is a considerable reduction  
in values $q_{\rm on,   min}$ and, respectively,  
$C_{\rm min}=2q_{\rm in} +q_{\rm on,   min}$   (Figs.~\ref{Charac_F_S_asym} (b, c)) in comparison with these values found in
 Sec.~\ref{Range_S}  (Fig.~\ref{Min_Capacity}).
Other peculiarities of the asymmetric F$\rightarrow$S transition have been found when $q_{\rm on}$ decreases:
(i)
The   value $R_{\rm OA}$     decreases strongly
(Fig.~\ref{Charac_F_S_asym} (b)). (ii) The discontinuity in the over-acceleration rate $R_{\rm OA}$ remains
until some inflow-rate denoted by $q^{\rm (1-lane)}_{\rm on,   max}$ that  slightly exceeds $q_{\rm on,   min}$
(Fig.~\ref{Charac_F_S_asym} (d)).  (iii) Although within a range  $q_{\rm on,   min}\leq q_{\rm on}<q^{\rm (1-lane)}_{\rm on,   max}$
free flow  is still metastable with respect to the asymmetric F$\rightarrow$S transition, nevertheless,
the discontinuity in the over-acceleration rate $R_{\rm OA}$ does not exist any more:
there is no lane-changing within the   inflow-rate range $q_{\rm on,   min}\leq q_{\rm on}<q^{\rm (1-lane)}_{\rm on,   max}$ at all.
To understand this result, we consider in Sec.~\ref{LWR-Sec}  automated-driving traffic on a single-lane road
with the same bottleneck.

\subsubsection{Over-acceleration in automated-driving traffic on single-lane road
 \label{LWR-Sec}}

After the asymmetric F$\rightarrow$S transition has occurred, {\it no}
effect of the bottleneck on the vehicle motion in the left lane is realized any more
(right column in Figs.~\ref{F_S_asym} (b--e));
therefore, each of the road lanes  could be considered
as two different (and not connected) single-lane roads.

\begin{figure}
\begin{center}
\includegraphics[width = 8 cm]{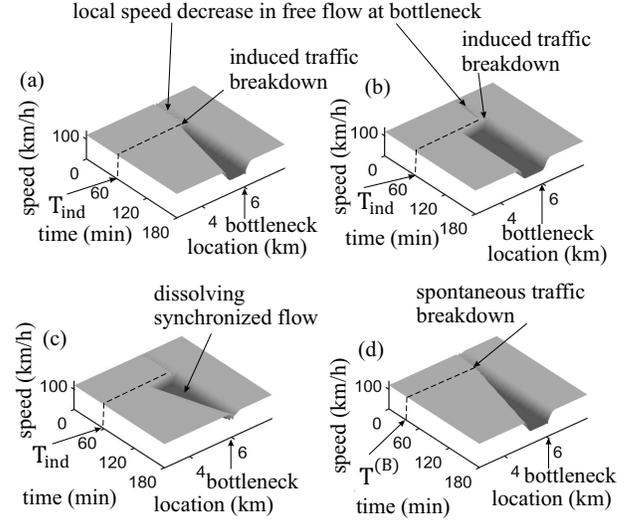}
\end{center}
\caption[]{Simulations of  
  F$\rightarrow$S transition  on single-lane road    
	with  bottleneck   model  of Sec.~\ref{Model_Sec}.
Speed in space and time   at different
  $q_{\rm on}$ at the same value $q_{\rm in}=$ 2571
 vehicles/h  as that in Fig.~\ref{Free_Flow_Bottl}:
(a) Induced traffic breakdown in metastable free flow at $q_{\rm on}=$ 365 vehicles/h, 
$T_{\rm ind}=$ 60 min, $\Delta q_{\rm on}=$ 135 vehicles/h, $\Delta t=$ 1 min;
(b) Induced traffic breakdown in metastable free flow at $q_{\rm on}=$ 360 vehicles/h, 
$T_{\rm ind}=$ 60 min, $\Delta q_{\rm on}=$ 540 vehicles/h, $\Delta t=$ 2 min;
(c) Dissolving synchronized flow at $q_{\rm on}=$ 350 vehicles/h, 
$T_{\rm ind}=$ 60 min, $\Delta q_{\rm on}=$ 550 vehicles/h, $\Delta t=$ 2 min;
(d) Time-delayed  spontaneous traffic breakdown at $q_{\rm on}=$ 366 vehicles/h, 
$T^{\rm (B)}=$ 30 min.    $q_{\rm on, min}=$ 360 vehicles/h, 
$q^{\rm (1-lane)}_{\rm on, max}=$ 365.2 vehicles/h.
Other model parameters are the same as those in Fig.~\ref{Free_Flow_Bottl}. 
}
\label{F_S_1-lane}
\end{figure} 

We have found that although no
R$\rightarrow$L lane-changing is possible on the single-lane road,
within the range  $q_{\rm on,   min}\leq q_{\rm on}<q^{\rm (1-lane)}_{\rm on,   max}$
free flow  is  indeed in a metastable state with respect to the   F$\rightarrow$S transition  at the bottleneck
  (Fig.~\ref{F_S_1-lane} (a--c)). The maximum on-ramp inflow-rate  $q^{\rm (1-lane)}_{\rm on,   max}$
	determines the maximum capacity of automated-driving traffic on the single-lane road with the bottleneck:
	$C_{\rm max}=q_{\rm in}+q^{\rm (1-lane)}_{\rm on,   max}$: At a given $q_{\rm in}$, when
	$q_{\rm on}>q^{\rm (1-lane)}_{\rm on,   max}$ after a time delay
	$T^{\rm (B)}$, which is a decreasing on-ramp inflow-rate function,
	the   F$\rightarrow$S transition occurs spontaneously  at the bottleneck (Fig.~\ref{F_S_1-lane} (d)).

Thus,
	the minimum on-ramp inflow-rate $q_{\rm on,   min}$ of   free flow   metastability
	with respect to the asymmetric F$\rightarrow$S transition  at the bottleneck on two-lane road is determined by
	the minimum on-ramp inflow-rate $q_{\rm on,   min}$ of  free flow metastability on   single-lane road
	with the same bottleneck. To explain this result, we should recall that  
 in  three-phase traffic theory~\cite{KernerBook,KernerBook2,KernerBook3}, the term {\it over-acceleration} determines driver acceleration behaviors associated with a time delay in
	 acceleration that causes
  free flow metastability
	with respect to an F$\rightarrow$S transition  at a bottleneck.  In Helly's model 
	(\ref{ACC_General}), (\ref{ACC_g_opt}) there is a time delay in
	 acceleration. For this reason, it is not surprising that
	Helly's model 
	(\ref{ACC_General}), (\ref{ACC_g_opt}) shows over-acceleration on the single-lane road.
	However, the effect of this over-acceleration is practically insignificant: the range of
	the free flow metastability on   single-lane road is only $q^{\rm (1-lane)}_{\rm on,   max}-q_{\rm on,   min}=$ 
	5 vehicles/h (Fig.~\ref{Charac_F_S_asym} (d))~\footnote{Note that for Helly's model 
	(\ref{ACC_General}), (\ref{ACC_g_opt}) with $\tau_{\rm d}=$ 1.5 s, 
	we have found $q^{\rm (1-lane)}_{\rm on,   max}-q_{\rm on,   min}=$ 
	10 vehicles/h.}.

\subsection{Effect of desired time headway of automated vehicles}

\begin{figure}
\begin{center}
\includegraphics[width = 8 cm]{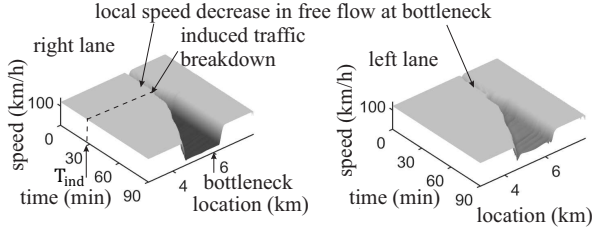}
\end{center}
\caption[]{Simulations of the effect of the increase in  desired time headway $\tau_{\rm d}$ of automated vehicles on
   nucleation features of F$\rightarrow$S transition with the model of Sec.~\ref{Model_Sec} 
	on two-lane road with bottleneck:
Speed in space and time in the  right lane (left) and in   left lane (right).    
Induced traffic breakdown  
under condition (\ref{range_formula}). $\tau_{\rm d}=$ 1.5 s with $K_{1}=0.3 \ s^{-2}$, $K_{2}=0.6 \ s^{-1}$ in 
(\ref{ACC_General}), (\ref{ACC_g_opt}), $\tau_{1}=\tau_{2}=$ 0.9 s in (\ref{g_prec_ACC}),
$q_{\rm in}=$ 1714
(vehicles/h)/lane,
  $q_{\rm on}=$ 710 vehicles/h,     $T_{\rm ind}=$ 30 min,
	$\Delta q_{\rm on}=$ 190 vehicles/h, $\Delta t=$ 2 min.
Other model parameters are the same as those in Fig.~\ref{Free_Flow_Bottl}. 
}
\label{1-5_F_S}
\end{figure}

The basic result about the metastability of free flow with respect to the F$\rightarrow$S transition
at the bottleneck remains under a wide range  of the desired time headway $\tau_{\rm d}$ of automated  vehicles.
However, as shown in Fig.~\ref{1-5_F_S}, the increase in $\tau_{\rm d}$ to 1.5 s leads to a considerable decrease in  
  the   flow rate $q_{\rm in}$ at which  the metastability of free flow is realized.

\section{Transitions between the three phases   in automated-driving vehicular traffic \label{WMJ_Sec}}

  Wide moving jams can emerge in synchronized flow. We have found that features of the jams  
are qualitatively almost the same as well-known
for human-driving traffic. Thus, we present a simplified   analysis 
of wide moving jams    for model of Sec.~\ref{Lane-asym_S}, when due to the use
of condition (\ref{speed_diff}) synchronized flow and 
wide moving jams can emerge in the right road lane only.

\begin{figure}
\begin{center}
\includegraphics[width = 8 cm]{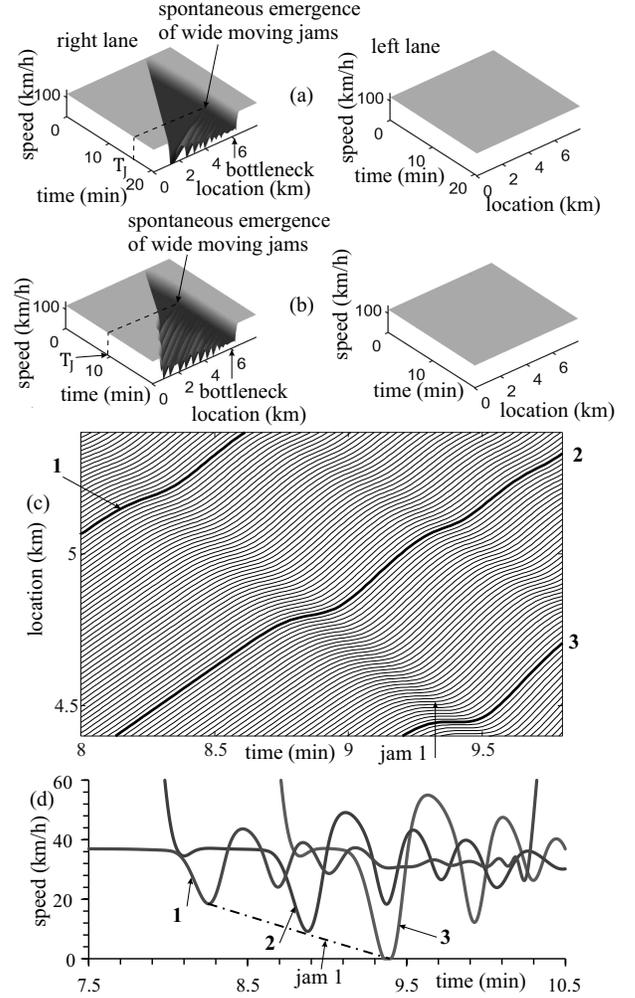}
\end{center}
\caption[]{Simulations of  spontaneous
  S$\rightarrow$J transition     in model of Sec.~\ref{Lane-asym_S} with the use of   (\ref{g_v_g_min}).
Speed in space and time in the right lane (left column) and in   left lane (right column) at different
  $q_{\rm on}$ at the same flow rate $q_{\rm in}=$ 2571
(vehicles/h)/lane as that in Fig.~\ref{Free_Flow_Bottl}:
(a) $q_{\rm on}=$ 900 vehicles/h; 
(b) $q_{\rm on}=$ 940 vehicles/h.  
(c) Vehicle trajectories in the right lane  for a part of (b).
 (d) Time-functions of speeds of vehicles 1, 2, and 3 shown in (c).
In (\ref{g_v_g_min}),
$v_{\rm min}=$ 36 km/h, $g_{\rm min}=$ 3 m.
Other model parameters are the same as those in Fig.~\ref{F_S_asym}. 
}
\label{S_J_asym}
\end{figure}

\begin{figure}
\begin{center}
\includegraphics[width = 8 cm]{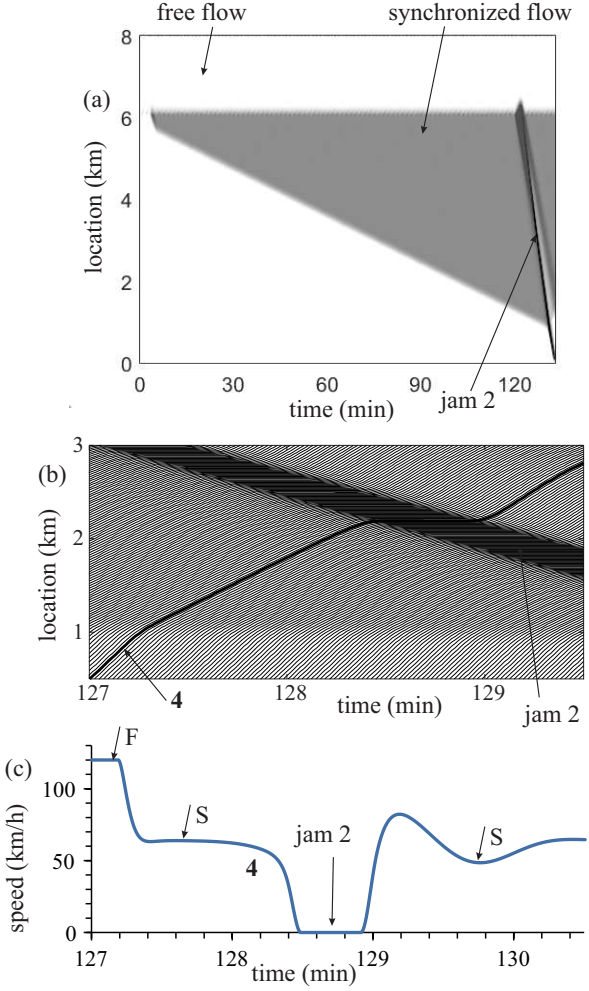}
\end{center}
\caption[]{Simulations of coexistence of the three phases F, S, and J
with the use of a sequence of induced F$\rightarrow$S and
  S$\rightarrow$J transitions     in model of Sec.~\ref{Lane-asym_S} 
	  with the use of   (\ref{g_v_g_min})
at the same flow rate $q_{\rm in}=$ 2571
(vehicles/h)/lane as that in Fig.~\ref{Free_Flow_Bottl}. (a) Speed data in space and time in the right lane
presented by regions with variable shades of gray [shades of gray
vary from white to black when the speed decreases from 120 km/h
(white) to 0 km/h (black)].
  $q_{\rm on}=$ 400 vehicles/h; for induced F$\rightarrow$S transition,  $T_{\rm ind}=$ 3 min,
	$\Delta q_{\rm on}=$ 500 vehicles/h, $\Delta t=$ 1 min; 
		for induced S$\rightarrow$J transition,  $T_{\rm ind}=$ 120 min,
	$\Delta q_{\rm on}=$ 800 vehicles/h, $\Delta t=$ 2 min. 
	(b) Vehicle trajectories  in the right lane  for a part of (a).
	(c) Time-function of   speed of vehicle 4 in (b).
	Wide moving jam (J) is marked by $\lq\lq$jam 2",
	F -- free flow, S -- synchronized flow.
Other model parameters are the same as those in Fig.~\ref{S_J_asym}. 
}
\label{S_J_asym_induced}
\end{figure}

For a study  of very low speed states in automated-driving vehicular traffic,  we should note that
in Eq.~(\ref{ACC_General}), when the speed $v\rightarrow 0$, the
optimal gap between vehicles $g_{\rm opt}$ (\ref{ACC_g_opt}) tends also to zero: $g_{\rm opt}\rightarrow 0$.
However, even when all vehicles are in standstill, the space gap between vehicles should be larger than zero. Therefore,
when the  vehicle speed  decreases below
  some low speed denoted by $v_{\rm min}$, in formula (\ref{ACC_g_opt}) we should add some additional space gap
denoted by $g_{\rm min}$ to which the space gap $g$ between automated vehicles tends when the speed $v\rightarrow 0$; therefore, formula (\ref{ACC_g_opt}) is replaced by a known formula
\begin{eqnarray}
\lefteqn{g_{\rm opt}=} \nonumber \\
& =\left\{\begin{array}{ll}
v\tau_{\rm d} &  \textrm{at $v \geq v_{\rm min}$,} \\
g_{\rm min}+v\left(\tau_{\rm d}
-\tau_{\rm min}\right) &  \textrm{at $v < v_{\rm min}$}, \\
\end{array} \right.
\label{g_v_g_min}
\end{eqnarray}
where $\tau_{\rm min}=g_{\rm min}/v_{\rm min}$;   $g_{\rm min}$  and $v_{\rm min}$ are constants.

\begin{figure}
\begin{center}
\includegraphics[width = 8 cm]{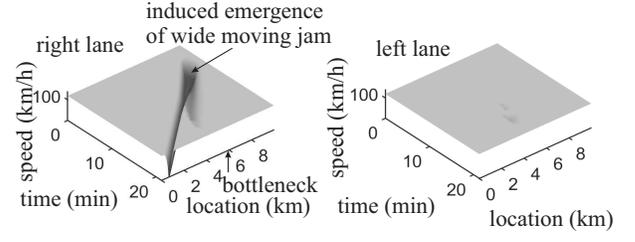}
\end{center}
\caption[]{Simulations of  induced
  F$\rightarrow$J transition     in model of Sec.~\ref{Lane-asym_S} with   condition (\ref{g_v_g_min}).  
Speed in space and time in the right lane (left column) and in   left lane (right column) 
 at  $q_{\rm in}=$ 2880 (vehicles/h)/lane and $q_{\rm on}=0$; for induced F$\rightarrow$J transition,  $T_{\rm ind}=$ 3 min,
	$\Delta q_{\rm on}=$ 1200 vehicles/h, $\Delta t=$ 2 min.  
Other model parameters are the same as those in Fig.~\ref{S_J_asym}. 
}
\label{F_J_asym}
\end{figure}

\begin{figure}
\begin{center}
\includegraphics[width = 7 cm]{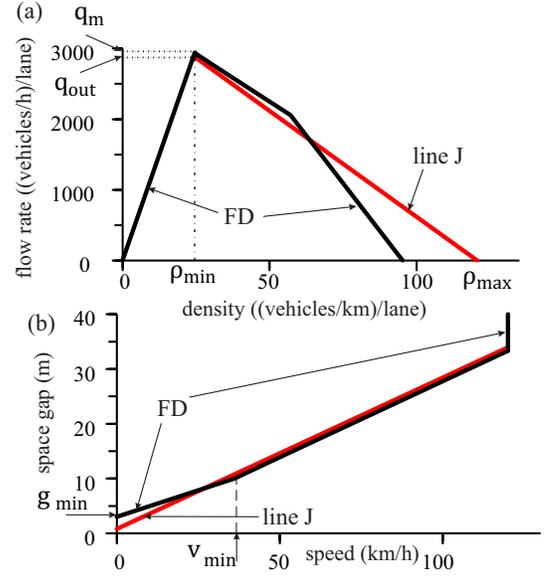}
\end{center}
\caption[]{ Fundamental diagrams   for hypothetical steady states  
(curves FD) and line J   in the flow--density plane   (a) and in the space-gap--speed plane (b).
Model of Sec.~\ref{Lane-asym_S} with the use of   (\ref{g_v_g_min}).
 The maximum flow rate on FD is $q_{\rm m}=3600/(\tau_{\rm d}+(d/v_{\rm free}))=$ 2939 (vehicles/h)/lane.
  Characteristics of line J   calculated for the wide moving jam in Fig.~\ref{F_J_asym} during the jam propagation in free flow
  are $v_{\rm g}=$ 30 km/h,
	 $q_{\rm out}=$ 2893 (vehicles/h)/lane, $\rho_{\rm min}=$ 24.5 (vehicles/km)/lane, $\rho_{\rm max}=$ 120.6 (vehicles/km)/lane.
Other model parameters are the same as those in Fig.~\ref{S_J_asym}. 
}
\label{Char_J_asym}
\end{figure}
  
	\begin{figure}
\begin{center}
\includegraphics[width = 8 cm]{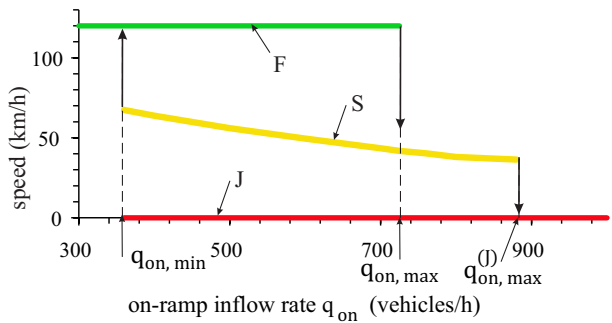}
\end{center}
\caption[]{Double Z (2Z)-characteristic for  transitions between the three
  phases F, S, and J:
On-ramp inflow-rate function of
average speed within the phases F, S, and J.
 Model of Sec.~\ref{Lane-asym_S} with the use of   (\ref{g_v_g_min}) at $q_{\rm in}=$ 2571
(vehicles/h)/lane   of Fig.~\ref{Free_Flow_Bottl}.  
	$q^{\rm (J)}_{\rm on, max}=$ 880 vehicles/h.
Other model parameters are the same as those in Figs.~\ref{Charac_F_S_asym}
and~\ref{S_J_asym}. 
}
\label{2Z_asym}
\end{figure}
	
We have found that in automated-driving traffic either a spontaneous S$\rightarrow$J transition
(Fig.~\ref{S_J_asym}) or
 induced S$\rightarrow$J transition  (Fig.~\ref{S_J_asym_induced}) can be realized.
Vehicle trajectories 1, 2, and 3 in Figs.~\ref{S_J_asym} (c, d) 
  show a typical example of  a time-development
of an emergent wide moving jam (marked by $\lq\lq$jam 1") during
the spontaneous S$\rightarrow$J transition. The dynamics of the induced S$\rightarrow$J transition
(Fig.~\ref{S_J_asym_induced} (a, b))
as well as a time-dependence of the speed of vehicle 4 propagating through
 the induced wide moving jam (marked by $\lq\lq$jam 2")  show   a possible coexistence of all three phases F, S, and J in automated-driving traffic
(Fig.~\ref{S_J_asym_induced} (c)) that is qualitatively very similar to that known for human-driving traffic. In  addition with 
S$\rightarrow$J transitions, a wide moving jam can be induced in free flow (induced F$\rightarrow$J transition)
(Fig.~\ref{F_J_asym}).

As in human-driving traffic, there are characteristics parameters of the downstream front
propagation of a wide moving jam in automated-driving traffic that do not depend on initial conditions.
The characteristic jam parameters presented by a   line J
in Fig.~\ref{Char_J_asym}  
are: (i) the velocity of the upstream propagation of the downstream jam front $v_{\rm g}$, (ii)
 the flow rate $q_{\rm out}$ and    (iii) the density
$\rho_{\rm min}$
in the jam outflow (when free flow is built in this jam outflow) as well as (iv)  the density within the jam
  $\rho_{\rm max}$.

States of free flow, synchronized flow, and wide moving jams build together
a   double-Z (2Z)-characteristic for   phase transitions in automated-driving traffic
(Fig.~\ref{2Z_asym}).
 At a given $q_{\rm in}$,
there is some maximum on-ramp inflow-rate $q_{\rm on}$ denoted by $q^{\rm (J)}_{\rm on, max}$ (Fig.~\ref{2Z_asym}).
The condition $q_{\rm on}=q^{\rm (J)}_{\rm on, max}$
  separates metastable synchronized flow  at
	$q_{\rm on}\leq q^{\rm (J)}_{\rm on, max}$ and unstable 
synchronized flow   at
	$q_{\rm on}> q^{\rm (J)}_{\rm on, max}$, when after a time delay $T_{\rm J}$
	a spontaneous S$\rightarrow$J transition is realized
	(Fig.~\ref{S_J_asym}).
The larger the difference $q^{\rm (J)}_{\rm on, max}-q_{\rm on}$, the shorter
the time delay $T_{\rm J}$ of the   S$\rightarrow$J transition
(Figs.~\ref{S_J_asym} (a, b)).

The 2Z-characteristic shows (Fig.~\ref{2Z_asym})
that any phase transitions between the three phases F, S, and J
are possible in a broad range of the flow rate in automated-driving vehicular traffic
on two-lane road at the bottleneck.

\section{Discussion \label{Dis_Sec}}
 
 We have shown that traffic  on a two-lane road with a bottleneck
that consists of 100$\%$  string-stable  automated  vehicles   moving in a road lane in accordance with
 the classical Helly's model~\cite{Helly_1959}
   is described in the framework  of the three-phase traffic theory in which
	traffic breakdown is an F$\rightarrow$S transition that exhibits the nucleation nature.
Does  this basic paper result remain when  vehicle platoons are 
  string-unstable  (Sec.~\ref{String-unstable_Sec}) or when
  a qualitatively different model for   automated-driving vehicles
is used (Sec.~\ref{TPACC_Sec})?

\subsection{F$\rightarrow$S transition at bottleneck in  automated-driving vehicular traffic under string-unstable conditions
\label{String-unstable_Sec}}

\begin{figure}
\begin{center}
\includegraphics[width = 7 cm]{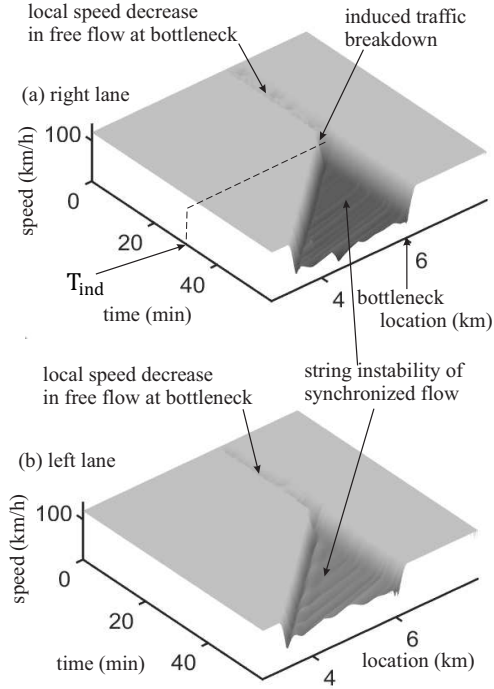}
\end{center}
\caption[]{Simulations of F$\rightarrow$S transition    on two-lane road with   bottleneck of model
of Sec.~\ref{Model_Sec} with the use of  
(\ref{g_v_g_min}), however, when condition (\ref{ACC_stability}) for string stability is  {\it not} satisfied:
Speed in space and time in the right lane (left) and in  left lane (right)
at $q_{\rm in}=$ 2571
(vehicles/h)/lane of Fig.~\ref{Free_Flow_Bottl}.    Induced F$\rightarrow$S transition
  simulated as   in Fig.~\ref{Induced_F_S_Bottl}.
 $\tau_{\rm d}=$ 1 s, $K_{1}=0.3 \ s^{-2}$, $K_{2}=0.75 \ s^{-1}$,
  $q_{\rm on}=$ 650 vehicles/h,     $T_{\rm ind}=$ 30 min,
	$\Delta q_{\rm on}=$ 250 vehicles/h, $\Delta t=$ 2 min.
Other model parameters are the same as those in Fig.~\ref{Free_Flow_Bottl}. 
}
\label{StringUnstable_F_S}
\end{figure}

\begin{figure}
\begin{center}
\includegraphics[width = 7 cm]{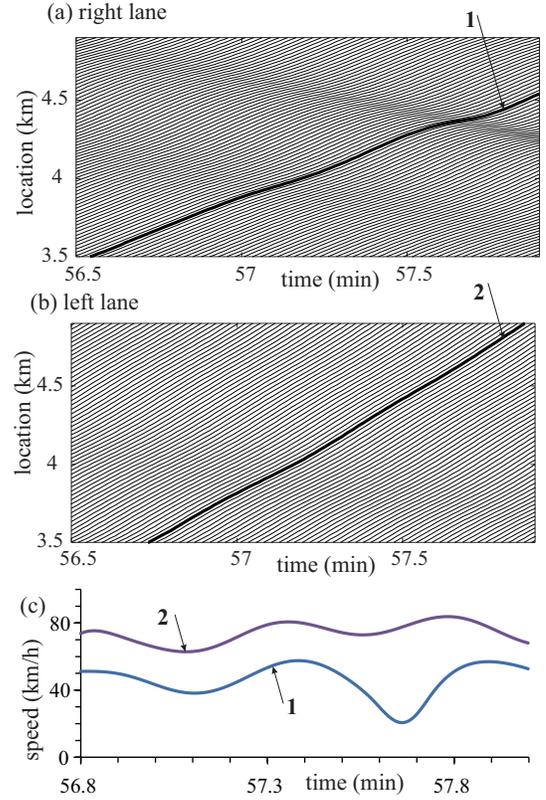}
\end{center}
\caption[]{Continuation of Fig.~\ref{StringUnstable_F_S}.
String instability of synchronized flow:
(a, b) Simulated vehicle trajectories  in synchronized flow
  in the right lane (a) and left lane (b) at time $t > T_{\rm ind} + \Delta t$.
 (c)
Time-functions of speeds  of vehicle 1 in the right lane   and vehicle 2
in the left lane maked by the same numbers
    in (a, b).  
}
\label{StringUnstable_F_S_tr}
\end{figure}

The basic paper result   about the nucleation nature of traffic breakdown
(F$\rightarrow$S transition) of the three-phase traffic theory
is valid for both string-stable 
and string-unstable automated-driving vehicular traffic (Fig.~\ref{StringUnstable_F_S}).  
In free flow, when the speed is equal to $v_{\rm free}$, the mean time headway between vehicles is longer than  
the desired value $\tau_{\rm d}$ in (\ref{ACC_General}), (\ref{ACC_g_opt}). Therefore,
no long enough vehicle platoons    in which automated vehicles moves at time headway
$\tau_{\rm d}$ can be built in free flow at the bottleneck: No string instability does   occur in free flow.
This explains why basic features of the free flow metastability with respect to the F$\rightarrow$S transition at the bottleneck remain qualitatively the same as those found in Secs.~\ref{Meta_S}--\ref{WMJ_Sec} for string-stable automated vehicles.

Contrary to free flow, in synchronized flow resulting from the F$\rightarrow$S transition at the bottleneck
very long vehicle platoons    in which automated vehicles moves at time headway
$\tau_{\rm d}$ can be built. For this reason, in synchronized flow the string instability 
  is realized (Fig.~\ref{StringUnstable_F_S_tr}). However,
a detailed study of the   development of the string instability in synchronized flow 
 that could be an interesting subject of scientific investigations
  is out of scope of this paper.

\subsection{Automated-driving   traffic based on three-phase adaptive cruise control (TPACC) \label{TPACC_Sec}}

\begin{figure}
\begin{center}
\includegraphics[width = 8 cm]{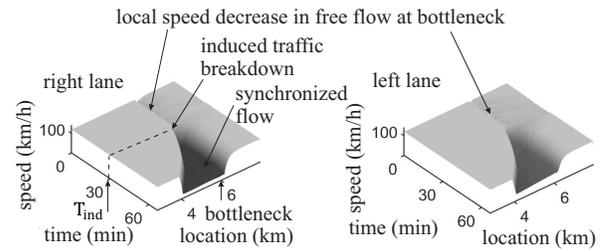}
\end{center}
\caption[]{Nucleation features of F$\rightarrow$S transition in automated-driving traffic consisting of 100$\%$  TPACC-vehicles
(\ref{TPACC_main5}) under the use of lane-changing and bottleneck models of Sec.~\ref{Model_Sec}:
Speed in space and time in the right lane (left column) and in left lane (right column).  
	$\tau_{\rm p}=$ 1.3 s, $\tau_{\rm G}=$ 1.4 s, $\tau_{\rm safe}=$ 1 s,  $K_{1}=0.3 \ s^{-2}$, $K_{\rm \Delta v}=K_{2}=0.6 \ s^{-1}$,
	$\tau_{1}=\tau_{2}=$ 0.5 s,
	$q_{\rm in}=$ 2000
(vehicles/h)/lane,
  $q_{\rm on}=$ 700 vehicles/h,     $T_{\rm ind}=$ 30 min,
	$\Delta q_{\rm on}=$ 200 vehicles/h, $\Delta t=$ 2 min.
Other parameters are the same as those in Fig.~\ref{Free_Flow_Bottl}. 
}
\label{TPACC_F_S}
\end{figure}

The basic result of the paper about the nucleation nature of traffic breakdown (F$\rightarrow$S transition)
 of the three-phase traffic theory remains when a qualitatively different model for   automated-driving vehicles
is used. In Fig.~\ref{TPACC_F_S},  
automated-driving   traffic based on three-phase adaptive cruise control (TPACC) is simulated.
The TPACC-model reads as 
follows~\cite{KernerTPACC,KernerTPACC_2}:  
\begin{eqnarray}
\lefteqn{a^{\rm (TPACC)}=} \nonumber \\
& =\left\{\begin{array}{ll}
K_{\rm \Delta v}\Delta v &   \textrm{at $g_{\rm safe} \leq g \leq G$}, \\ 
K_{1}(g-g_{\rm opt})+K_{2}\Delta v  & \textrm{at $g< g_{\rm safe}$ or $g> G$ },    
\end{array} \right. 
\label{TPACC_main5} 
\end{eqnarray}
where   $K_{\rm \Delta v}$ is a constant dynamic coefficient   ($K_{\rm \Delta v}>0$),
$g_{\rm opt}$ is given by   (\ref{g_v_g_min}) when in this formula     $\tau_{\rm d}$ is replaced by
model parameter $\tau_{\rm p}$    that satisfies condition 
$\tau_{\rm p}<\tau_{\rm G}$, $\tau_{\rm G}$ is a synchronization space gap;   $g_{\rm safe}=v \tau_{\rm safe}$, 
$\tau_{\rm safe}$ is a safe time headway.
In contrast with the    model (\ref{ACC_General}), (\ref{g_v_g_min}),
in the TPACC-model  (\ref{TPACC_main5})
 there is an indifference zone for car-following when time headway is  between  
 $\tau_{\rm safe}$ and   $\tau_{\rm G}$, i.e., there is
 no fixed desired time headway between vehicles in TPACC-vehicle platoons.
 For this reason, as shown in~\cite{KernerTPACC}, there is no string-instability in TPACC-vehicle platoons.
 
Simulations show that   nucleation features of the F$\rightarrow$S transition
 in automated-driving     based on the TPACC-model (\ref{TPACC_main5})
are qualitatively the same as those found in Secs.~\ref{Meta_S}--\ref{WMJ_Sec} for string-stable automated vehicular traffic with the use of Helly's model
(\ref{ACC_General}), (\ref{ACC_g_opt}). However,
there are some qualitative differences in synchronized flow behavior  
caused by  the indifference zone for car-following in the TPACC-model (\ref{TPACC_main5}).
For example, while the velocity of the upstream synchronized flow front for Helly's model
(\ref{ACC_General}), (\ref{ACC_g_opt}) 
is almost time-independent (Fig.~\ref{Induced_F_S_Bottl}), this velocity can depend on time   
in the TPACC-model (\ref{TPACC_main5}) 
(Fig.~\ref{TPACC_F_S}).
 A more detailed consideration of   three-phase traffic theory for automated-driving traffic 
based on the TPACC-model that could be an interesting subject of scientific investigations is out of scope of this paper.

\subsection{Conclusions}

1. The nucleation nature of traffic breakdown (F$\rightarrow$S transition) at a highway bottleneck, which is the basic feature 
of the three-phase traffic theory for human-driving traffic,
has been revealed for    vehicular traffic consisting of 100$\%$ of  automated-driving vehicles
moving on a two-lane road with an on-ramp bottleneck. As long as   lane changing in free flow
ensures a distribution of  on-ramp inflow
  between road lanes, 
this basic result remains in a broad range of model parameters of automated-driving vehicles.  

2. We have found that there is a discontinuity in the rate of lane-changing   from the right lane (neighborhood lane to on-ramp) to the left lane (passing lane)
(denoted as R$\rightarrow$L lane-changing). In its turn, this causes the discontinuity in the  over-acceleration rate: The rate of over-acceleration in free flow   is larger
than it is in synchronized flow.

3. The cause of the nucleation nature of traffic breakdown (F$\rightarrow$S transition) in automated-driving vehicular traffic at a bottleneck is the discontinuity in the  over-acceleration rate  together with the spatiotemporal competition between over-acceleration and speed adaptation. 
A larger rate of over-acceleration 
		in free flow causes the maintenance of free flow at the bottleneck;
		contrarily, a lower rate of over-acceleration 
		in synchronized flow causes the maintenance of synchronized flow at the bottleneck.

4. Through the spatiotemporal competition between over-acceleration and speed adaptation caused by lane-changing, at any time instant
there is a range of highway capacities between some minimum and maximum capacities; within the capacity range, an
F$\rightarrow$S transition can be induced; however, when the maximum capacity is exceeded, then
after some time-delay a spontaneous
  F$\rightarrow$S transition  occurs at the bottleneck.
All three-phases F (free flow), synchronized flow (S), and wide moving jam (J) can coexist each other
in automated-driving traffic. A diverse variety of  
phase transitions, which can occur between the phases F, S, and J,     determine the spatiotemporal dynamics of
  automated-driving vehicular traffic.

5.  The discontinuous character of over-acceleration  caused by lane-changing  is   the universal physical feature  of vehicular traffic. The three-phase traffic theory
 is the framework for both human-driving and automated-driving vehicular traffic. Therefore,
we can assume that the three-phase traffic theory is also the framework for
a mixed traffic consisting of a random distribution of human-driving and automated-driving  vehicles.
Three-phase traffic theory of mixed traffic that is out of the scope of this paper could be a very interesting task for further traffic studies.

\begin{acknowledgments} 

I would like to thank Sergey Klenov for help in simulations and useful suggestions.
I thank our partners for their support in the project $\lq\lq$LUKAS -- Lokales Umfeldmodell f{\"u}r das Kooperative, Automatisierte Fahren in komplexen Verkehrssituationen" funded by the German Federal Ministry for Economic Affairs and Climate Action.
\end{acknowledgments}

\end{document}